\normalfont
\documentclass[prb,twocolumn,showpacs,preprintnumbers,amsmath,amssymb]{revtex4}
\usepackage{float}
\usepackage[pdftex]{graphicx}
\usepackage{dcolumn}
\usepackage{bm}
\usepackage[usenames]{color}
\usepackage{color}
\usepackage{multirow}

\allowdisplaybreaks[1]

\begin{document}


\title{Correlation effects on ground-state properties of ternary Heusler alloys: first-principles study}

\author{V.D.\ Buchelnikov$^{1,2}$}
\author{V.V.\ Sokolovskiy$^{1,2}$}
\author{O.N.\ Miroshkina$^{1}$}
\author{M.A.\ Zagrebin$^{1,2,3}$}

\author{J.\ Nokelainen$^{4}$}
\author{A.\ Pulkkinen$^{4}$}
\author{B.\ Barbiellini$^{4,5}$}
\author{E.\ L\"{a}hderanta$^{4}$}
\affiliation{$^{1}$Faculty of Physics, Chelyabinsk State University, 454001 Chelyabinsk, Russia}
\affiliation{$^{2}$National University of Science and Technology "MISiS", 119049 Moscow, Russia}
\affiliation{$^{3}$National Research South Ural State University, 454080 Chelyabinsk, Russia}
\affiliation{$^{4}$Lappeenranta University of Technology, FI-53851 Lappeenranta, Finland}
\affiliation{$^{5}$Department of Physics, Northeastern University, Boston, MA 02115, USA}

\date{\today}

\begin{abstract}
The strongly constrained and appropriately normed
(SCAN) semi-local functional for exchange-correlation is deployed to study the ground-state properties of ternary Heusler alloys transforming martensitically. The calculations are performed for ferromagnetic, ferrimagnetic, and antiferromagnetic phases. 
Comparisons between SCAN and generalized gradient approximation (GGA) are discussed. We find that SCAN yields smaller lattice parameters and higher magnetic moments compared to the GGA corresponding values for both austenite and martensite phases. Furthermore, in the case of ferromagnetic and non-magnetic Heusler compounds, 
GGA and SCAN display similar trends in the total energy as a function of lattice constant and tetragonal ratio. However, for some ferrimagnetic Mn-rich Heusler compounds, different magnetic ground states are found within GGA and SCAN. 

%

\end{abstract}


\maketitle

\section{Introduction}

Nowadays, the density functional theory (DFT) has become an accurate and efficient first-principles approach to investigate broad areas of physics, chemistry, and materials sciences with the aim of  understanding and predicting complex and novel systems at the nanoscale\cite{Kohn-1999,Becke-2014,Martin-2004}. The main merits of DFT consist of a reduction in the number of degrees of freedom by replacing 3$N$ coordinates with only three coordinates of the electron density $n$($\bf{r}$) and of the possibility to include the electron correlation beyond Hartree-Fock theory \cite{Hohenberg-1964,Kohn-1965,Kohn-1999}. The accuracy and efficiency of DFT is provided by the choice of the exchange-correlation (XC) functional, which includes many-body and quantum effects. 
In the "Jacob's ladder" scheme~\cite{Perdew-2001}, there are several rungs associated with consecutive improvement of correlation to achieve an arbitrary level of accuracy. However, higher rungs can become computationally challenging. The first rung is the local density approximation (LDA)\cite{Vosko-1980,Perdew-1981,Perdew-1986}, where the XC functional depends only on the local density $n$($\bf{r}$). The second one is the generalized gradient approximation (GGA) with no free parameters \cite{Perdew-1991,Burke-1997}, which depends on the local density gradient $\bigtriangledown n$($\bf{r}$). The most successful and widely used GGA parametrization has been proposed by Perdew, Burke, and Ernzerhof (PBE)\cite{Perdew-1996}. The third rung, the meta-GGA functional~\cite{Perdew-1999,Tao-2003}, includes also a dependence on the kinetic energy density $\tau$($\bf r$). The recently developed meta-GGA functional, called SCAN (strongly constrained and appropriately normed) \cite{Sun-2015}, 
has been found to perform better than GGA for calculations of
several systems with various types of bonding: intermediate-range van-der-Waals interactions~\cite{Sun-2016a} (right ordering of 7 polymorphs of H$_2$O ice), ionic bonding~\cite{Kitchaev-2016} (energetic ordering of 6 polymorphs of MnO$_2$), covalent and metallic bonds~\cite{Sun-2016b} (Si under different phases), lattice constants of 2D materials~\cite{Buda-2017}, and highly correlated materials~\cite{Lane-2018,Furness-2018,Zhang-2018} (La$_2$CuO$_4$, Sr-doped La$_2$CuO$_4$, and YBa$_2$Cu$_3$O$_{6+x}$). 
An extensive benchmark of SCAN has been performed recently by Isaacs and Wolverton \cite{Isaacs-2018} for a group of nearly 1000 crystalline compounds and compared to available experimental data. They found that SCAN provides more accurate crystal volumes and improved band gaps as compared to PBE.
However, Ekholm \textit{et al}. \cite{Ekholm-2018} shown that SCAN seems to improve the structural properties of bcc-Fe but does not give the overall improvement for itinerant ferromagnets. Similar conclusions were reached by Fu and Singh~\cite{Fu-2018} in the study of Fe and steel.  
Studies focusing on SCAN benchmarks in Heusler alloys have not been reported earlier except the work of Isaacs and Wolverton~\cite{Isaacs-2018} reporting about Ni$_2$XAl (X = Ti, Hf, and Nb) and Fe$_2$NiAl.

The full-Heusler alloys are of high experimental and theoretical interest due to the unique properties such as shape memory effect, superelasticity and superplasticity, giant magnetocaloric effect, giant magnetoresistance and magnetostrain~\cite{Vasil'ev-2003,Buchelnikov-2006,Entel-2006,Entel-2008,Bozhko-1998,  Buchelnikov-2008,Planes-2009,Aksoy-2009,Ye-2010,Graf-2010,Entel-2011,Entel-2012a,Entel-2014,Felser-2015,Prudnikov-2010,Granovskii-2012,Khovaylo-2013}, which makes them good candidates to be used in various technological applications. It is worth noting that the strong competition between FM and AFM interactions is a peculiarity of Mn-rich Heusler alloys, where the Mn-Mn exchange interactions reveal the long-range oscillatory behavior of the Rudermann-Kittel-Kasuya-Yosida (RKKY) interaction \cite{Kasuya-1974,Goncalves-1972,Shi-1994,Sasioglu-2004,Sasioglu-2008,Entel-2011}. 
Nowadays, there is a good amount of knowledge on ternary and quaternary Heusler alloys accumulated by various \textit{ab initio} studies~\cite{Entel-2006,Entel-2011,Entel-2012a,Sasioglu-2004,Sasioglu-2008,Li-2011,Chakrabarti-2013,
Entel-2014,Zeleny-2014,Xiao-2012,Xiao-2014,Comtesse-2014,Sokolovskiy-2015,Li-2015,Roy-2016,Dutta-2016,Matsushita-2017,Neibecker-2017,
Opeil-2008,Kundu-2017,Godlevsky-2001,Ayuela-2002,Ayuela-1999,Kart-2008,Ziewert-Dis,Hsu-2002,Kulkova-2004,Gupta-2014,Himmetoglu-2012,Hasnip-2013, Sokolovskiy-2014jpd,  Buchelnikov-2018,  Sokolovskiy-2019,  Buchelnikov-2019}, where 
considerable efforts were devoted to investigate the effect of the XC potential within LDA or GGA on the ground state properties. In general, GGA compared to LDA leads to more accurate phase diagrams for magnetic materials \cite{Bernardo-1990} but it is also  worthwhile to study corrections beyond the GGA scheme~\cite{Himmetoglu-2012,Hasnip-2013}.


In this work, we report the impact of SCAN corrections on the ground state properties of ternary intermetallics such as FM Ni$_{2+x}$Mn$_{1-x}$Ga and Fe$_2$Ni$_{1+x}$Ga$_{1-x}$, ferrimagnetic (FIM) Ni$_2$Mn$_{1+x}$(Ga,~Sn)$_{1-x}$, and non-magnetic Fe$_2$VAl. 
This collection of ternary Heusler alloys provides an overview of various FM, AFM, and FIM interactions among transition metal atoms such as V, Mn, Fe, and Ni. We also consider
 the binary compound NiMn since it is an end point for the phase diagram of the Mn-rich Ni$_{2}$Mn$_{1+x}$Z$_{1-x}$ family~\cite{Entel-2011}.
To investigate the effect of corrections beyond GGA, PBE and SCAN calculations are compared. The outline of the paper is as follows.  Section II is devoted to the description of the computational methods used in the simulations. Section~III presents results of the main ground state properties, total energy curves and density of states. Important discussions and conclusions are presented in Section~IV.

\section{Details of calculations}

DFT calculations were performed using the 
 plane-wave basis set and the projector augmented wave  (PAW)  method
as implemented in Vienna \textit{ab initio} simulation package (VASP) \cite{Kresse-1996,Kresse-1999}. GGA and meta-GGA XC functionals using PBE and SCAN parameterizations were employed.
The PAW pseudopotentials were used with the following atomic configurations: Mn (3$p^6$3$d^6$4$s^1$), Ni~(3$p^6$3$d^9$4$s^1$), Fe (3$p^6$3$d^7$4$s^1$), V~(3$p^6$3$d^4$4$s^1$), Al~(3$s^2$3$p^1$), Sn (4$d^{10}$5$s^2$5$p^2$), and Ga~(3$d^{10}$4$s^2$4$p^1$). 
 For all calculations, the plane wave basis kinetic energy cut-off of 550~eV was applied, whereas 
the kinetic energy cut-off for the augmented charge was chosen as 800~eV.
The uniform Monkhorst-Pack mesh of $8\times8\times8$ 
  $k$-points 
    together with a Gaussian broadening of 0.2 eV were used
 to integrate the Brillouin zone with the second order Methfessel-Paxton method. The calculations were converged with the energy accuracy of~$10^{-7}$~eV/atom. 

\begin{figure}[b!!] 
%
\centerline{
\includegraphics[width=8.50cm,clip]{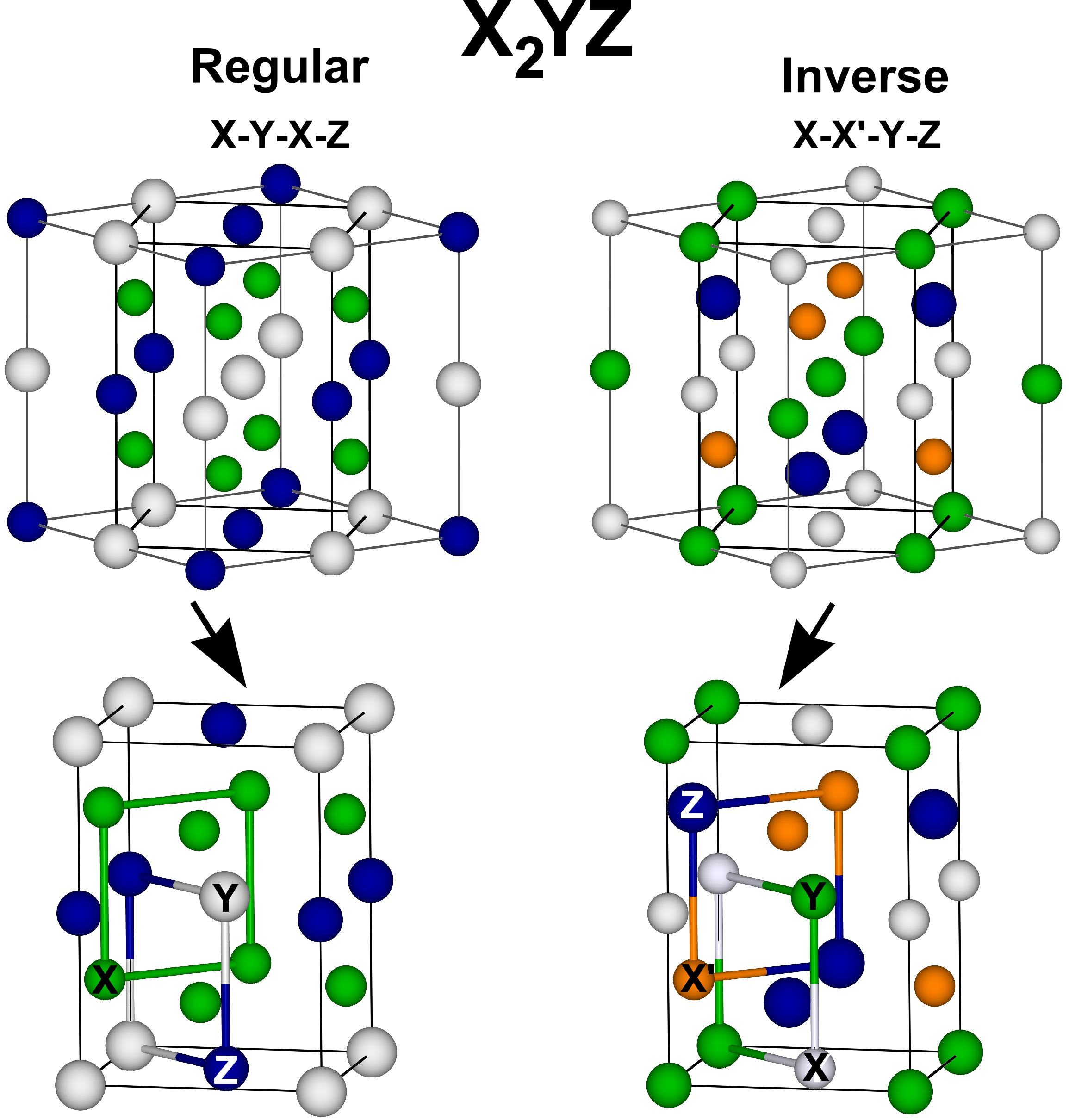}}
\caption{(Color online) 
Crystal structure of the full-Heusler X$_2$YZ alloys with regular Cu$_2$MnAl-type and inverse 
Hg$_2$TiCu-type. Bold solid lines highlight the tetragonal cells as shown in bottom panel. In our calculations, we considered two 8-atom supercells based on corresponding tetragonal cells.   
}
\label{figure-1}
%
\end{figure}

\begin{figure*}[!htb] 
\centerline{
\includegraphics[width=7.5cm]{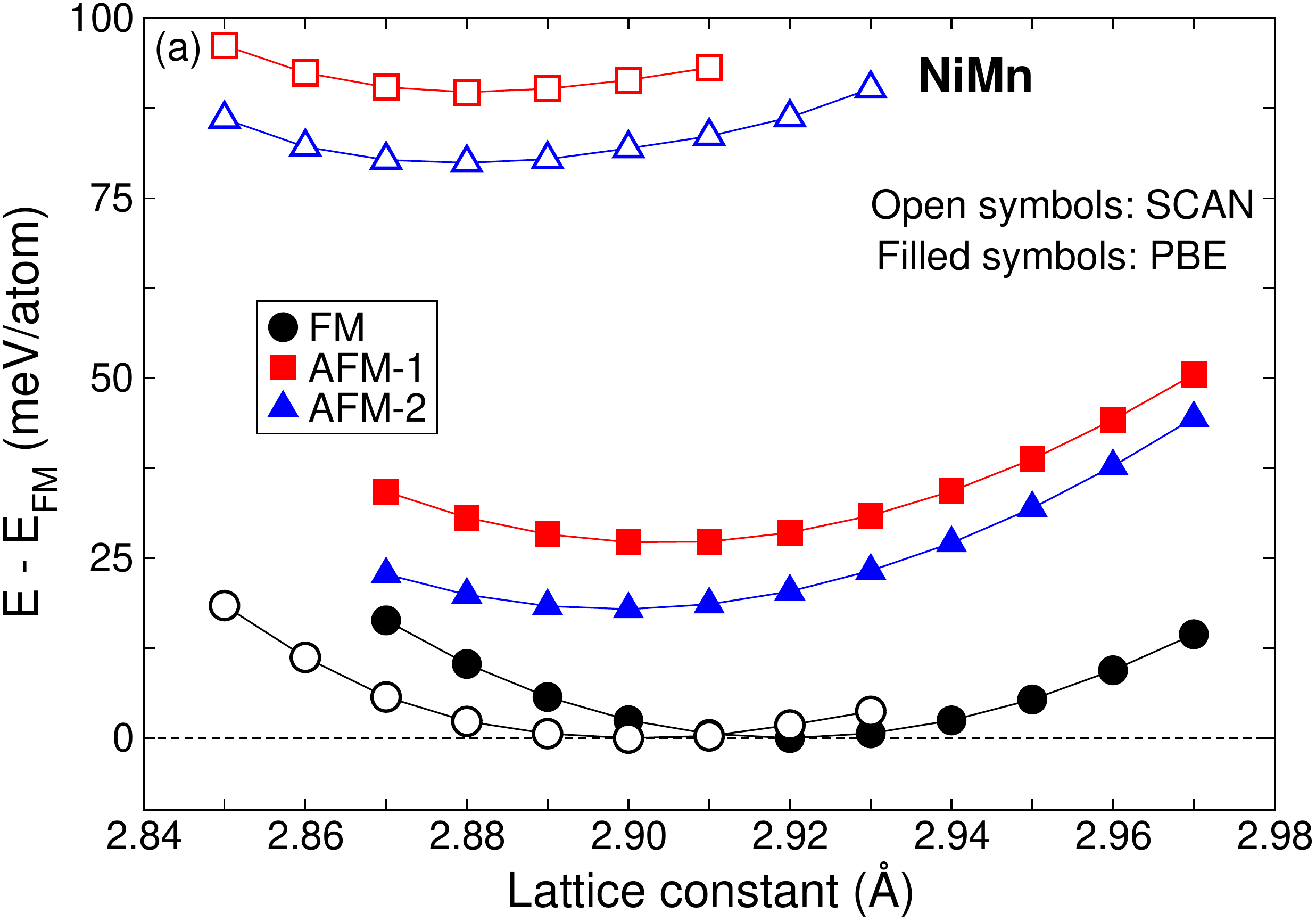}
 \hspace*{0.5cm}
\includegraphics[width=7.5cm]{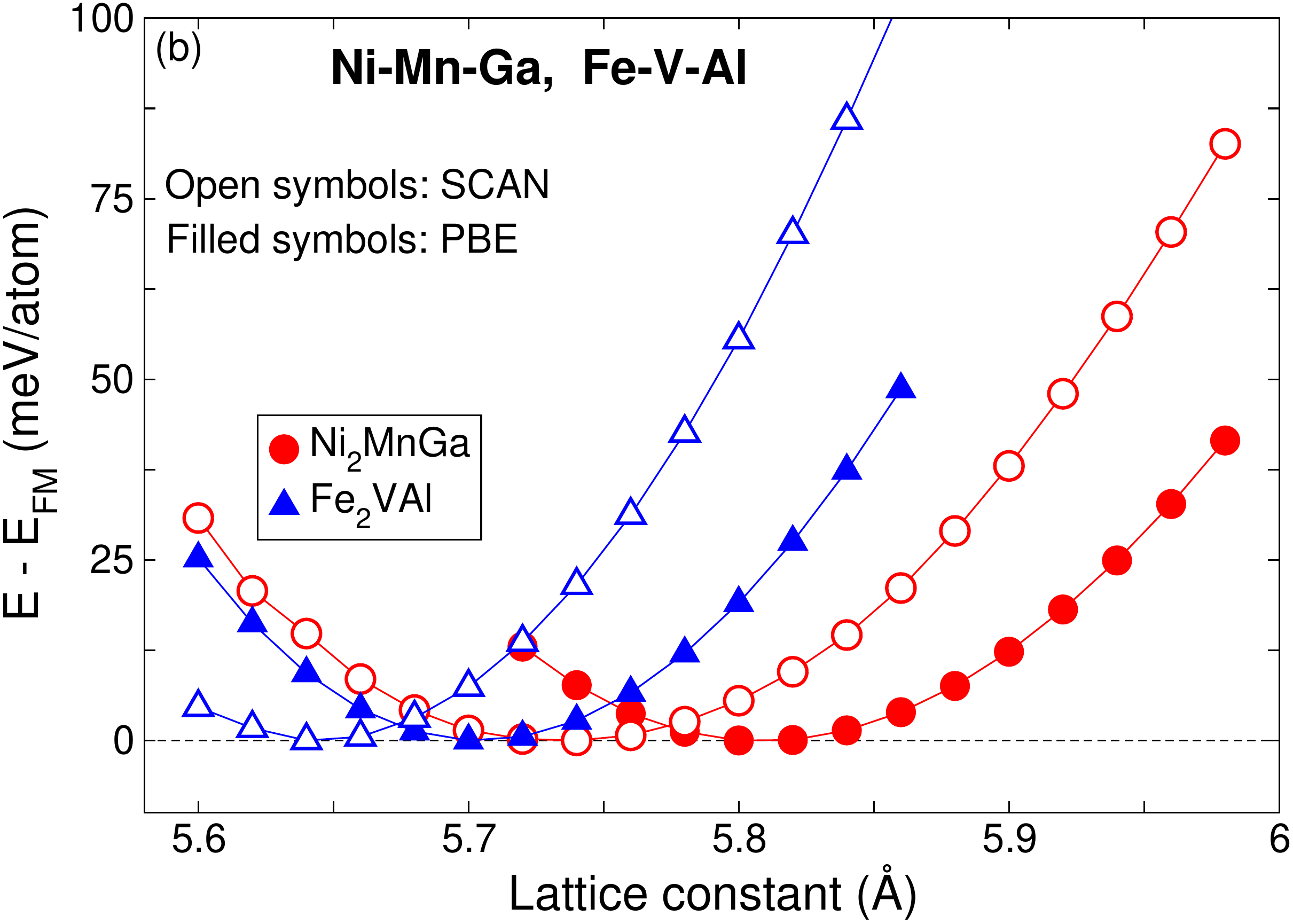}
           }   
           \centerline{
\includegraphics[width=7.5cm]{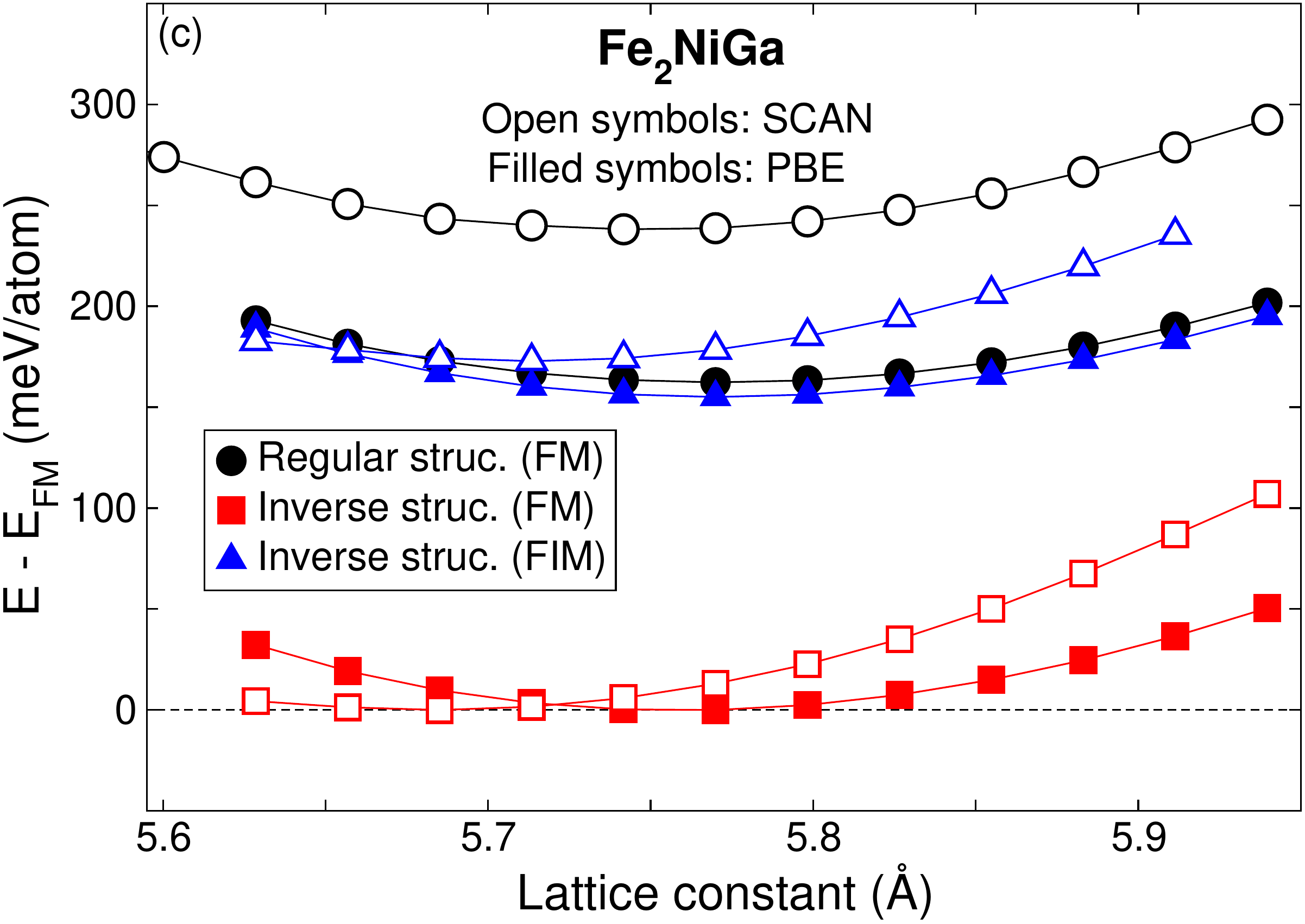}
 \hspace*{0.5cm}
\includegraphics[width=7.5cm]{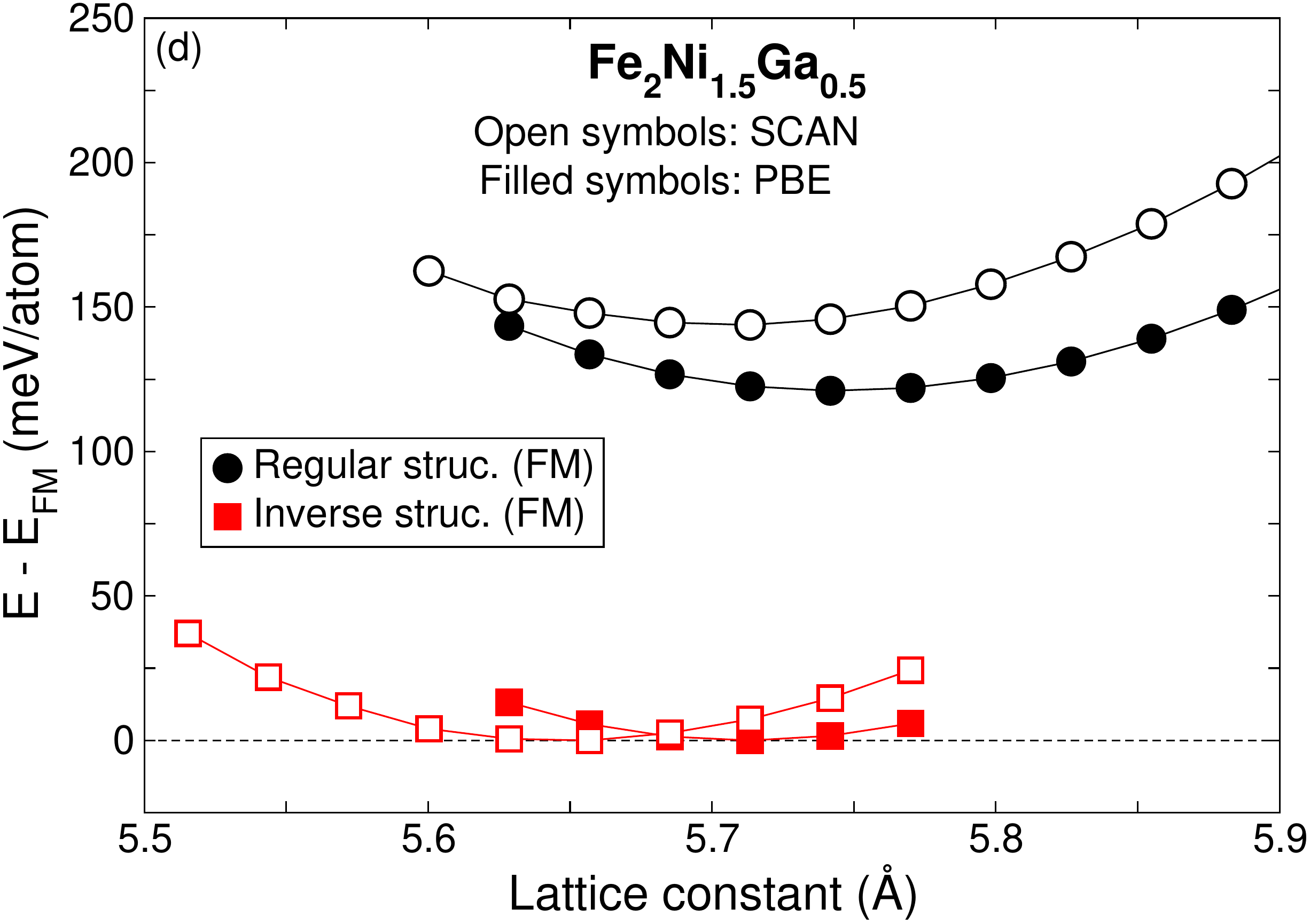}
           }  
           \centerline{
\includegraphics[width=7.5cm]{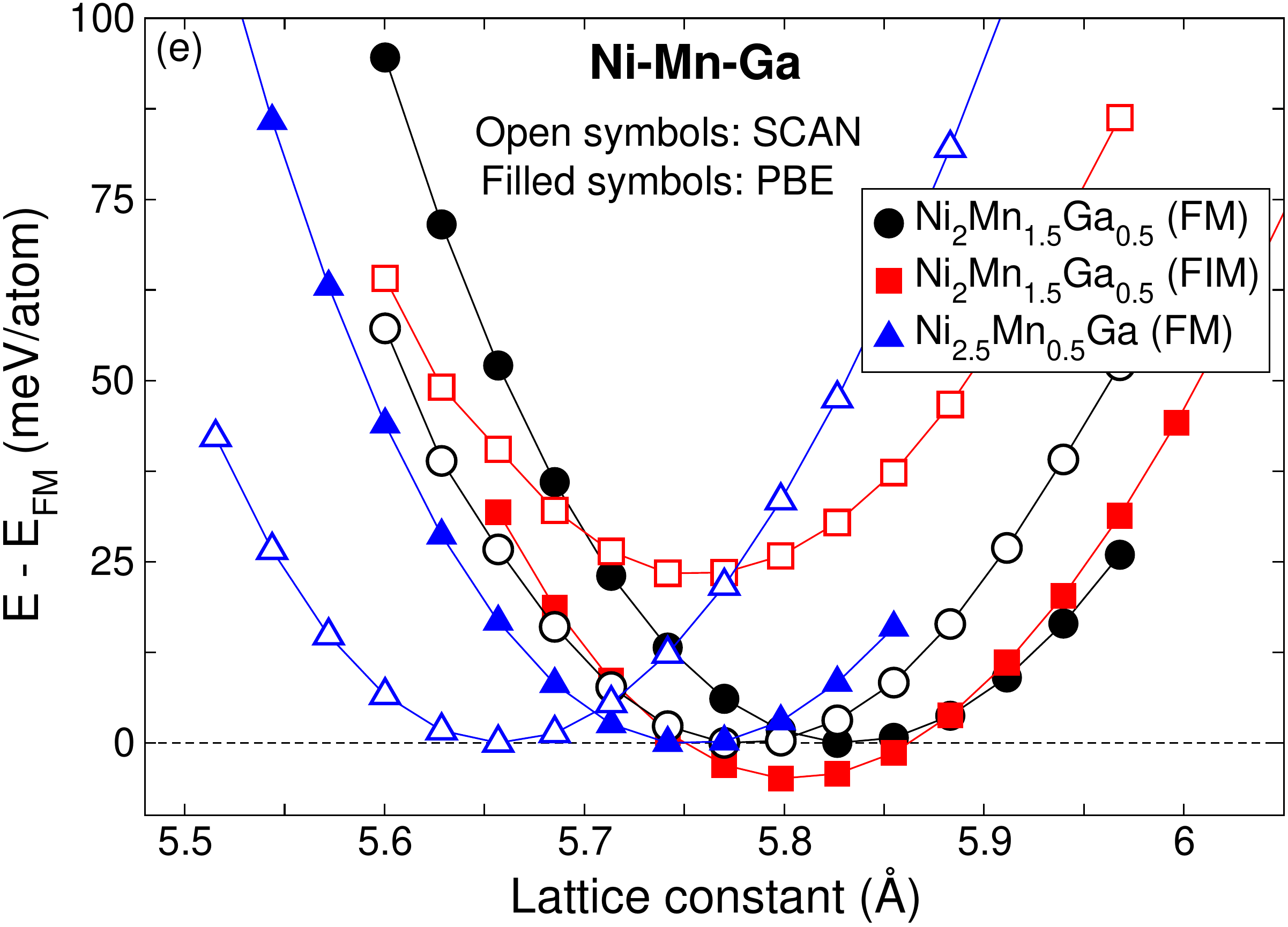}
 \hspace*{0.5cm}
\includegraphics[width=7.5cm]{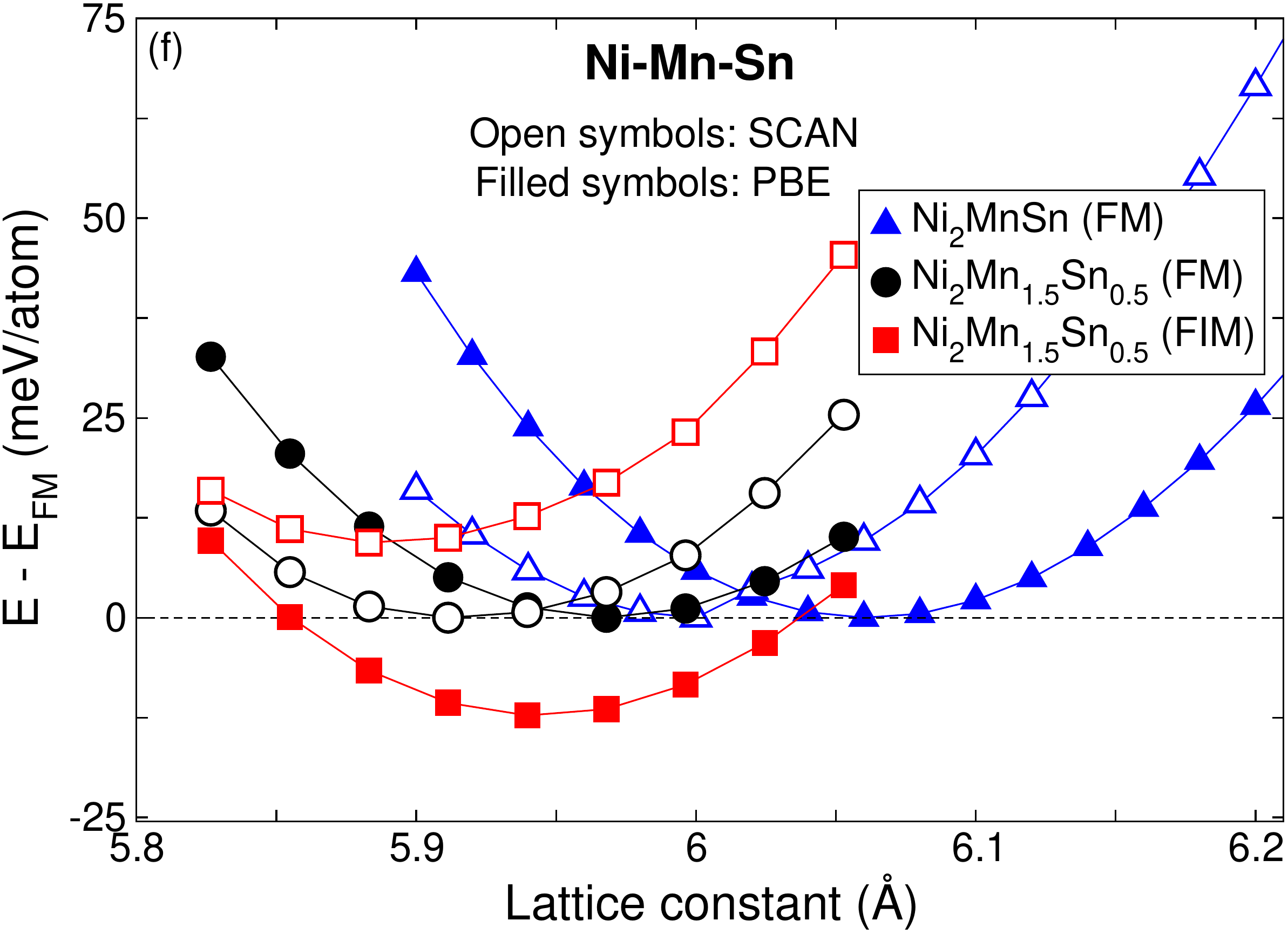}
           }  
\caption{(Color online) 
  The calculated total energy differences relative to the FM state as a function of lattice constant for (a)~NiMn, (b)
 Ni$_2$MnGa and Fe$_2$VAl, (c, d) Fe$_2$Ni$_{1+x}$Ga$_{1-x}$ ($x = 0$ and 0.5), (e) Ni$_{2.5}$Mn$_{0.5}$Ga and Ni$_2$Mn$_{1.5}$Ga$_{0.5}$, (f)~Ni$_2$Mn$_{1+x}$Sn$_{1-x}$ ($x = 0$ and 0.5). Results are presented for different magnetic configurations. In the case of Fe-Ni-Ga, the energy difference is plotted relative to the FM state of the inverse Heusler structure.
} 
\label{figure-2} 
%
\end{figure*}

In order to reduce computational costs, 
 all compositions were modeled using the 8-atom supercell of Heusler alloys with regular X$_2$YZ structure (space group $Fm\bar{3}m$ No.~225, prototype Cu$_2$MnAl) and inverse (XX')YZ structure (space group $F\bar{4}3m$ No.~216, prototype Hg$_2$TiCu) as shown in Fig.~\ref{figure-1}.
 In the case of regular Heusler structure, the unit cell contains four atoms as basis wherein two X atoms occupy 
8$c$ (1/4, 1/4, 1/4) and (3/4, 3/4, 3/4), whereas Z and Y atoms are placed at  
4$a$ (0, 0, 0) and 4$b$ (1/2, 1/2, 1/2) Wyckoff positions, respectively. In the inverse Heusler structure, two X atoms are situated at two distinct crystallographic sites 4$a$ (0, 0, 0) and 4$c$ (1/4, 1/4, 1/4) as well as Y and Z atoms are located at 4$b$ (1/2, 1/2, 1/2) and 4$d$ (3/4, 3/4, 3/4) sites. According to Graf \textit{et al}.~\cite{Graf-2010},  X$_2$YZ compound crystallizes in the inverse Heusler structure with space group $F\bar{4}3m$ on condition if the valence electrons of Y are more than those of X, otherwise it crystallizes in the regular Heusler structure ($Fm\bar{3}m$). 
Thus, calculations were performed for Ni$_{2+x}$Mn$_{1-x}$Ga, Ni$_2$Mn$_{1+x}$(Ga,~Sn)$_{1-x}$, and Fe$_2$VAl using the regular structure while for Fe$_2$Ni$_{1+x}$Ga$_{1-x}$, regular and inverse structures were employed.
For Ni$_2$Mn$_{1+x}$(Ga, Sn)$_{1-x}$ and Fe$_2$NiGa, FM and FIM orders were considered: FM corresponds to the parallel alignment of Ni, Mn, and Fe magnetic moments, while 
 for Mn-excess compounds, FIM corresponds to the antiparallel orientation of the magnetic moment on the Mn excess atoms (occupying Ga and Sn sites). For Fe$_2$Ni$_{1+x}$Ga$_{1-x}$, the magnetic moments of Fe atoms placed at the 4$a$ sites are opposed to those of Fe atoms situated at the 4$c$ sites.     

\section{Results and discussion}
\subsection{Ground-state crystal structure}

\begin{table*}[!htb] 
\caption{The equilibrium lattice constants $a_0$, bulk moduli $B$, and formation energies per formula unit $\Delta E_{\mathrm{form}}$ in the cubic structure of considered alloys calculated with PBE and SCAN. Also the other calculated (calc.) and experimental (exp.) data are given. Notice, the reference theoretical parameters have been calculated without meta-GGA. 
}
\begin{ruledtabular}
\begin{tabular}{ccc cc cc cc cc}
System& \multicolumn{2} {c}{$a_0$ [\AA]}& \multicolumn{2} {c}{$B$ [GPa]}& \multicolumn{2} {c}{ $\Delta E_{\mathrm{form}}$ [eV/f.u.]} &$a_0$ [\AA] & $B$ [GPa]  \\
      &  PBE& SCAN & PBE & SCAN  & PBE & SCAN & (Ref.) & (Ref.)  \\ \hline
    NiMn (bcc)      & 2.92    & 2.901   &  149.4  &  134.7 &    0.141  & -0.425 &  2.94 (calc)\cite{Godlevsky-2001},  2.97 (calc.)\cite{Egorushkin-1983}   &155 (calc.)\cite{Godlevsky-2001} \\
          &    &   &  &   &   & & 2.93 (calc.)\cite{Entel-2011}, 2.917 (calc.)\cite{Busgen-2004}   &  &   &  \\
Ni$_2$MnGa &  5.809   & 5.737   & 154.4 & 153.1 &   -0.645  & -1.583   & 5.81 (calc.)\cite{Ayuela-1999}, 5.806(calc.)\cite{Entel-2006}   & 156 (calc.)\cite{Ayuela-1999} \\
   &    &   &  &   &   & & 5.812 (calc.)\cite{Kart-2008}, 5.822 (calc.)\cite{Entel-2011}   & 155 (calc.)\cite{Kart-2008} &   &  \\
    &    &   &  &   &   & & 5.825 (exp.)\cite{Webster-1984}, 5.822 (exp.)\cite{Cakir-2013} &  146 (exp.)\cite{Worgull-1996} &  \\

Ni$_{2.5}$Mn$_{0.5}$Ga    &  5.754   & 5.668 & 167.9 & 206.5  & -0.5   &   -1.100      &   5.811 (calc.)\cite{Entel-2011}   &  \\
Ni$_2$Mn$_{1.5}$Ga$_{0.5}$  & 5.805  & 5.781  &  148.6 &  159.5    &  -0.142        & -1.158    &  5.81 (calc.)\cite{Ziewert-Dis}    \\

Ni$_2$MnSn  &  6.06        & 5.99&   140.7  &   159.5 &    -0.149  & -0.832    &   6.059 (calc.)\cite{Ayuela-1999}, 6.06 (calc.) \cite{Entel-2011} & 140 (calc.)\cite{Ayuela-1999}   \\
 &    &   &  &   &   & & 6.057 (calc.)\cite{Entel-2006}, 6.046 (exp.)\cite{Krenke-2005}   & 146 (calc.)\cite{Li-2013} \\

Ni$_2$Mn$_{1.5}$Sn$_{0.5}$     & 5.944    & 5.92  & 140.9  &  145.4 &  0.113&   -0.686      &  5.95 (calc.)\cite{Entel-2011}, 6.0 (calc.)\cite{Xiao-2012}   &  \\
     Fe$_2$VAl  &  5.704 & 5.644 & 218.5 & 252.6  &  -1.691     &  -1.699   &  5.712 (calc.) \cite{Hsu-2002}, 5.76 (exp.)\cite{Hsu-2002} & 212 (calc.) \cite{Hsu-2002}     \\
 Fe$_2$NiGa   & 5.759  & 5.682  & 172.7 & 179.6   &  -0.426   & -1.136   & 5.78 (calc.) \cite{Kulkova-2004}, 5.76 (calc.)\cite{Gupta-2014} & 146 (calc.) \cite{Kulkova-2004}  \\
 &    &   &  &   &   & &  5.77 (calc.) \cite{Matsushita-2017}, 5.81 (exp.) \cite{Gasi-2013}   & 174 (calc.) \cite{Gupta-2014}\\
 Fe$_2$Ni$_{1.5}$Ga$_{0.5}$      &  5.712   &  5.648   & 179.1   & 186.2   &  0.124  & -0.600   &  \\
 
\end{tabular}
\end{ruledtabular}
\label{table-1}
\end{table*}

We perform geometry optimization calculations to
investigate the ground-state properties for the cubic austenite structure using both PBE and SCAN. 
Figure~\ref{figure-2} shows the total energy as function of the lattice constant
for binary NiMn and ternary Ni$_{2+x}$Mn$_{1-x}$Ga ($x = 0$ and 0.5),  
Ni$_2$Mn$_{1+x}$(Ga, Sn)$_{1-x}$ ($x = 0$ and 0.5), Fe$_2$Ni$_{1+x}$Ga$_{1-x}$ ($x = 0$ and 0.5), and Fe$_2$VAl. For the binary NiMn system shown in Fig.~\ref{figure-2}(a), the calculations were performed for FM and AFM (AFM-1 and AFM-2) configurations~\cite{Entel-2011}. AFM-1 and AFM-2 correspond to the layered and staggered magnetic structure, respectively. Both PBE and SCAN give FM as the ground state for bcc-NiMn. Moreover, results for AFM-1 and AFM-2 follow similar trends for PBE and SCAN but the SCAN equilibrium lattice constants are smaller than PBE one. 

We next consider the results for stoichiometric and off-stoichiometric Ni- and Fe-based Heusler alloys (Figs.~\ref{figure-2}(b)-(f)). The SCAN equilibrium lattice parameters are smaller than the PBE ones in all these cases. PBE and SCAN predict correctly the inverse Heusler structure\cite{Gasi-2013} for Fe$_2$NiGa (Fig.~\ref{figure-2}(c)) with FM order. Similar results are obtained for Fe$_2$Ni$_{1.5}$Ga$_{0.5}$ as shown in Fig.~\ref{figure-2}(d).
For Ni$_2$Mn$_{1.5}$(Ga, Sn)$_{0.5}$ (see Figs.~\ref{figure-2}(e, f)), PBE yields the FIM ground state of~L2$_1$-cubic structure, which agrees with earlier results \cite{Entel-2014,Xiao-2012,Xiao-2014}, whereas SCAN finds the FM order more favorable than the FIM one. This disagreement might be explained as follows. Since SCAN gives smaller lattice constant, the neighbor Mn-Mn distance is also reduced leading to a modification of the RKKY interactions. 
However, one should note that RKKY interaction is only important in an asymptotic limit, for the magnetic ground state the closer neighbor couplings are more relevant. Thus, in this case, the Bethe-Slater (BS) curve~\cite{Cardias-2017} should provide a more pertinent tool to rationalize the magnetism of compound Mn compounds (see also Ni$_2$MnAl~\cite{Galanakis-2011, Simon-2015}). Since Mn is situated at a point of the BS curve where the AFM and FM orders are near in energy, a critical parameter controlling the type of magnetic order is provided by the next neighbor distance between Mn atoms. Consequently, larger separations result in ferromagnetism and smaller distances are connected to AFM order. 

\begin{figure*}[!htb] 
\centerline{
\includegraphics[width=7.5cm]{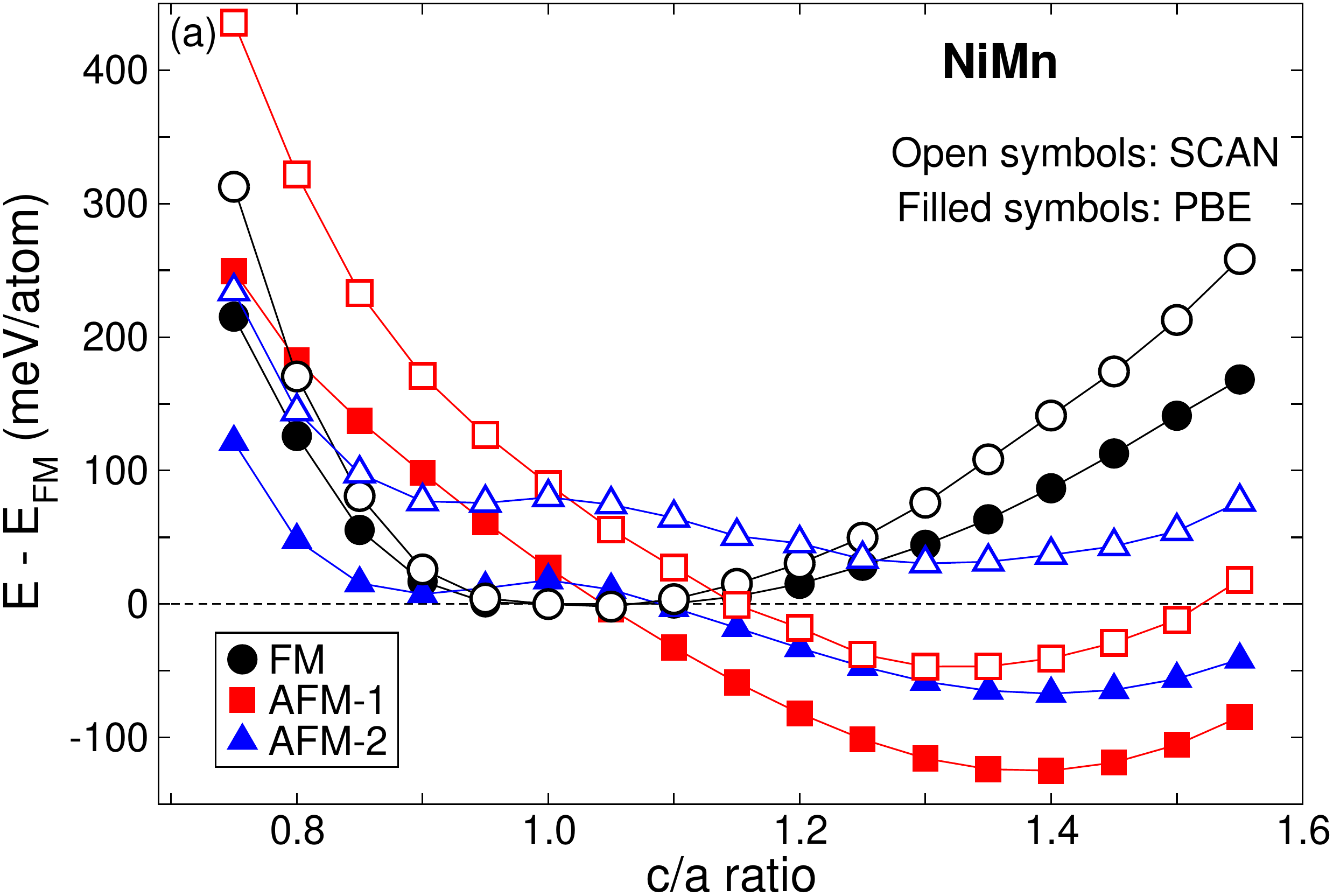}
 \hspace*{0.5cm}
\includegraphics[width=7.5cm]{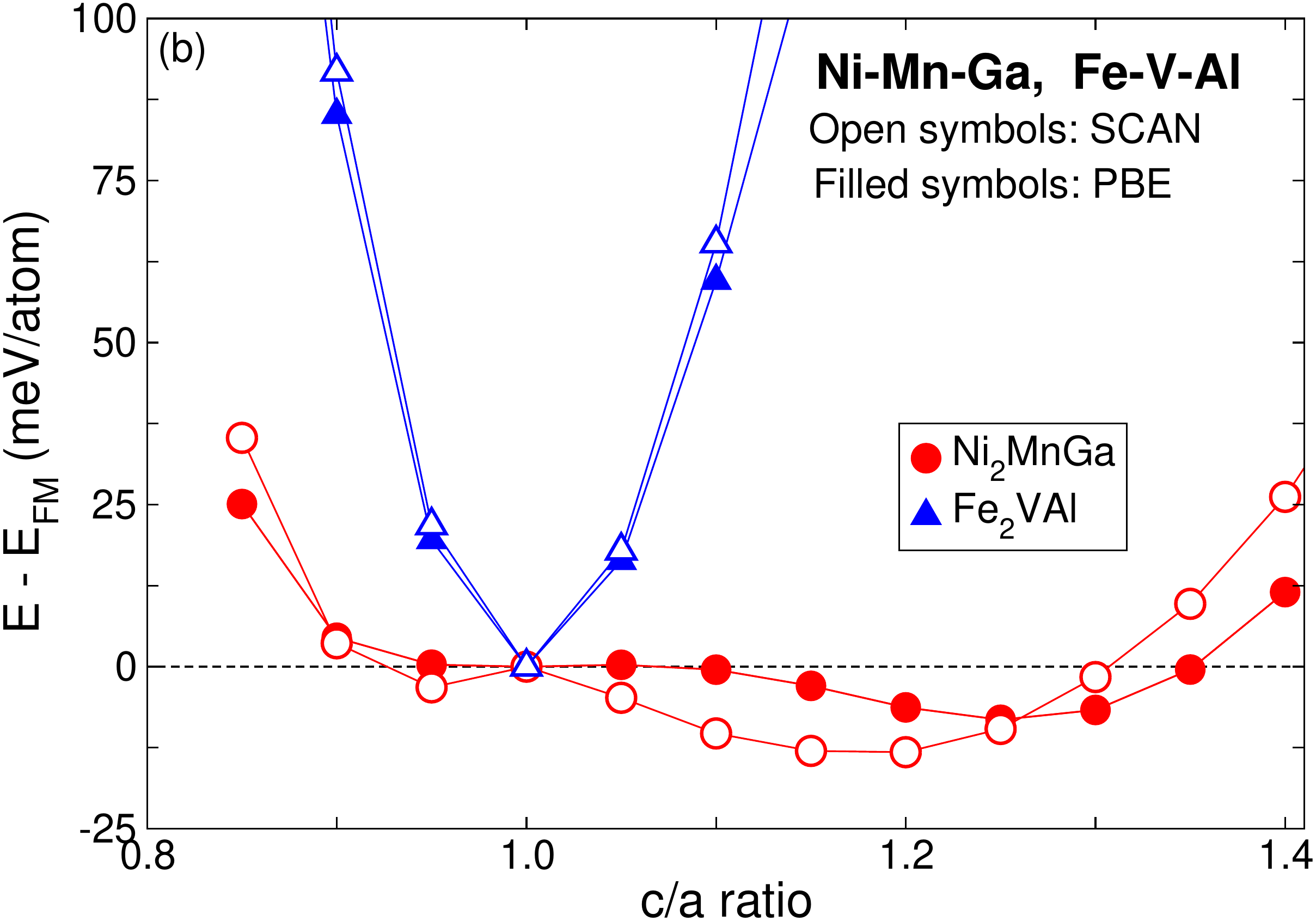}
           }   
           \centerline{
\includegraphics[width=7.5cm]{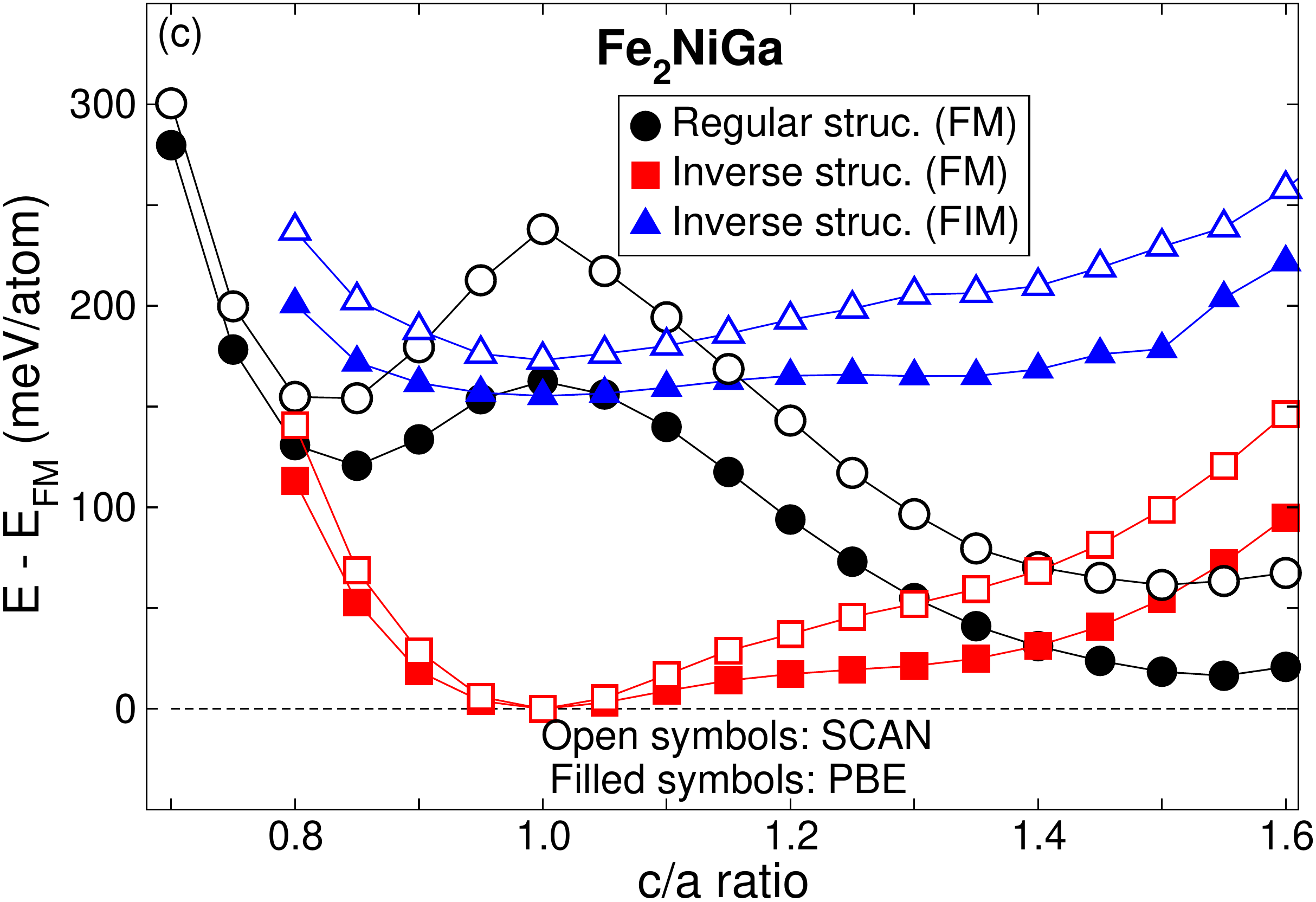}
 \hspace*{0.5cm}
\includegraphics[width=7.5cm]{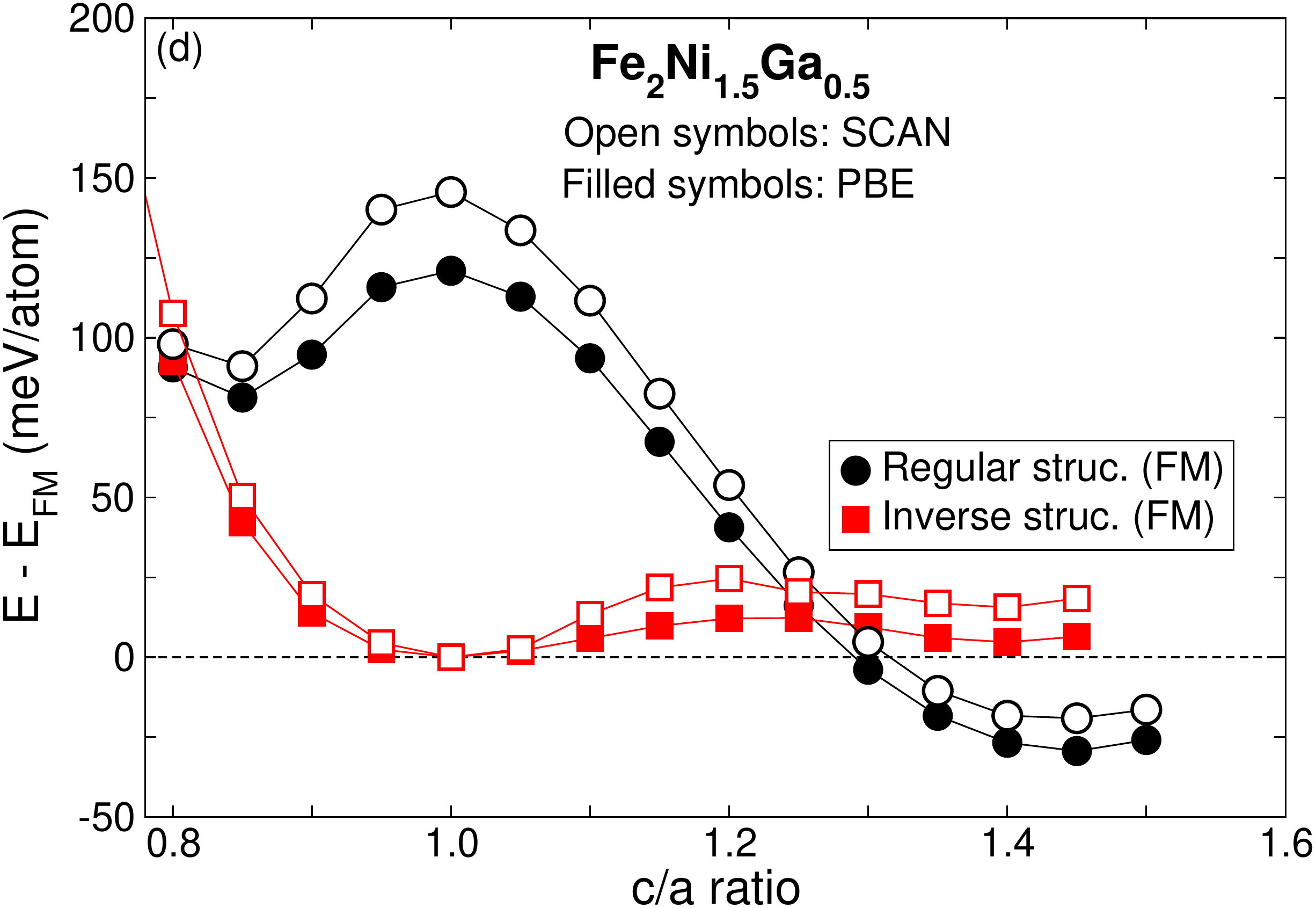}
           }  
           \centerline{
\includegraphics[width=7.5cm]{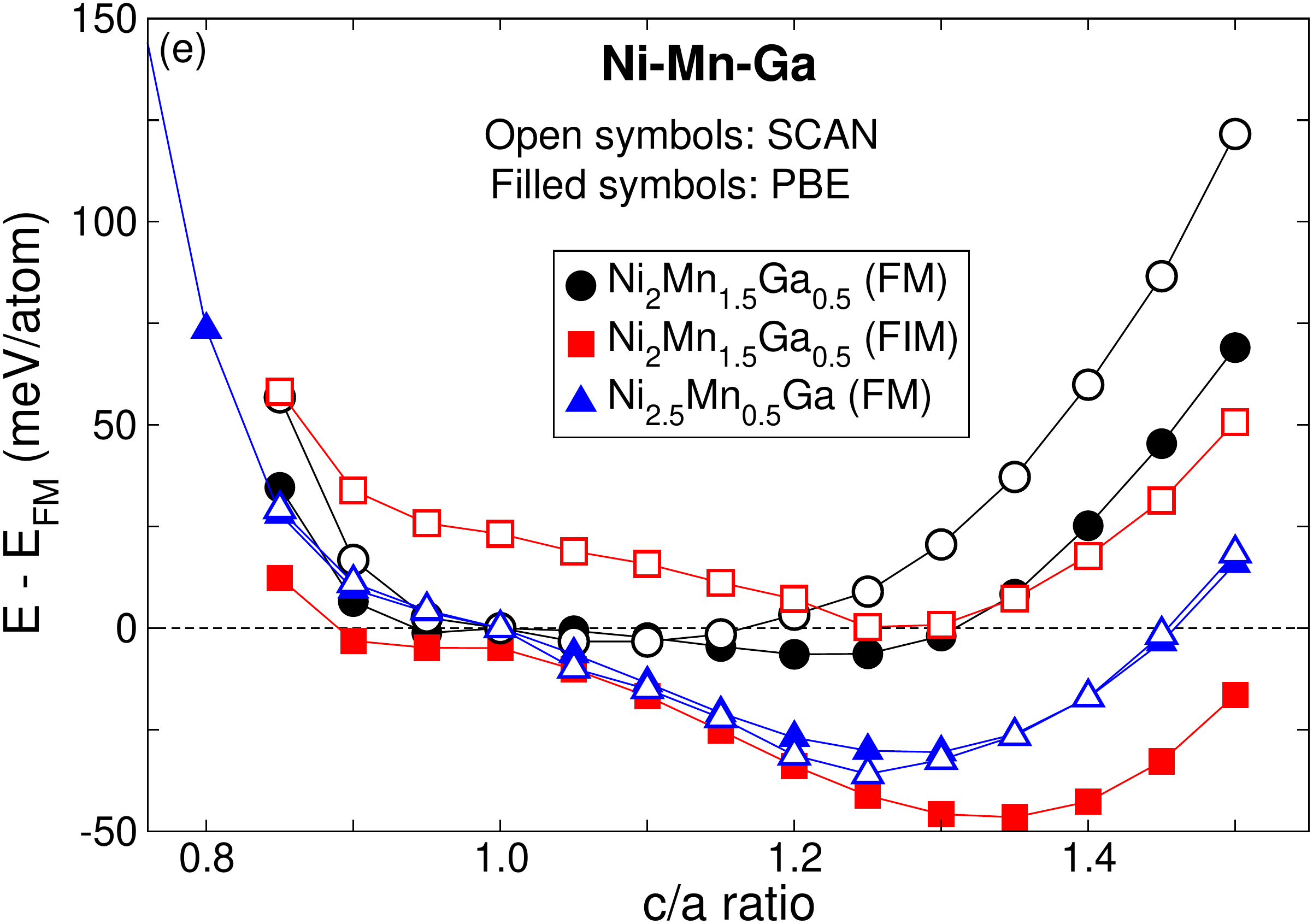}
 \hspace*{0.5cm}
\includegraphics[width=7.5cm]{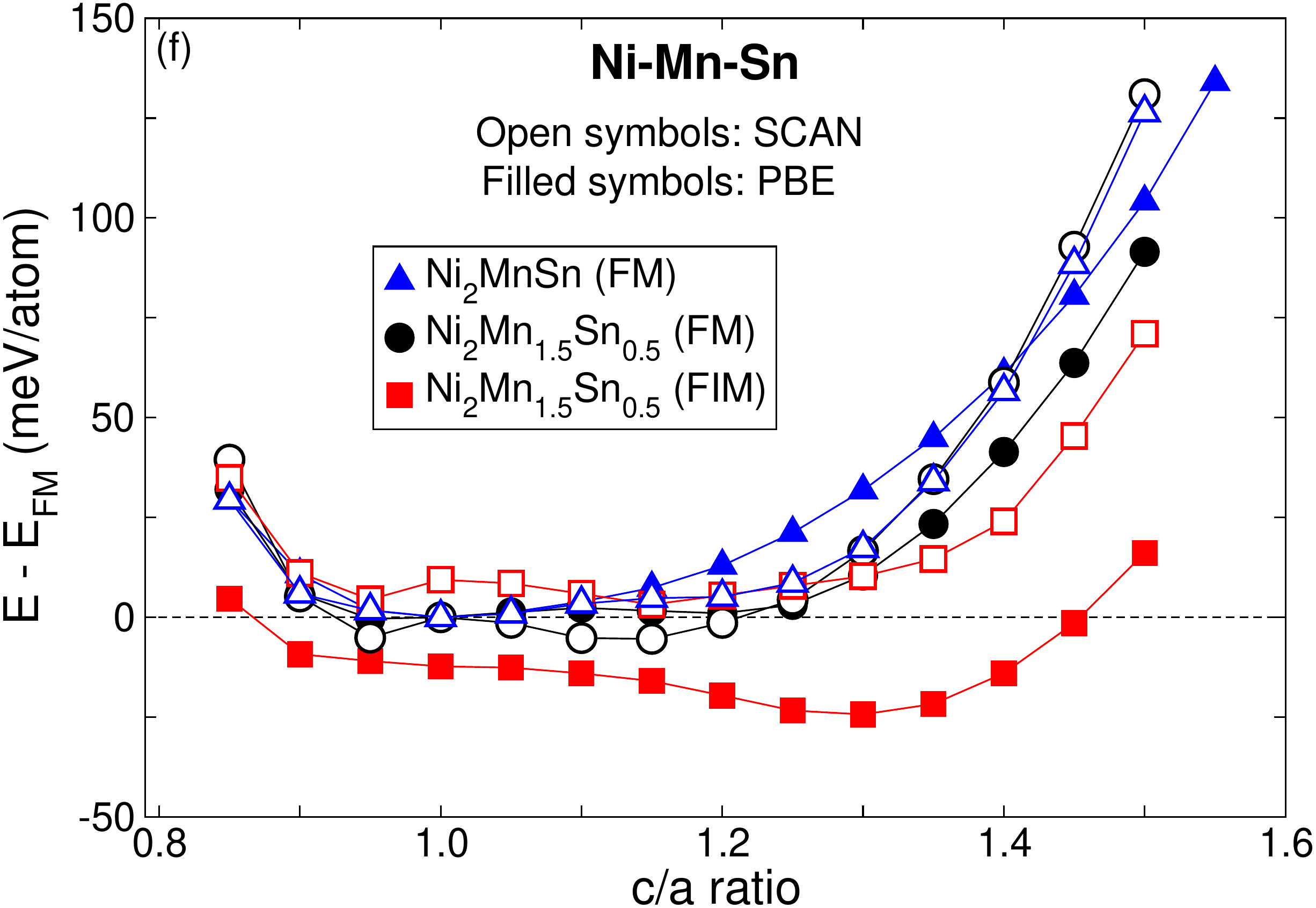}
           }  
\caption{(Color online) 
  The calculated total energy differences relative to the FM state as a function of tetragonal ratio $c/a$ for (a)~NiMn, (b)
 Ni$_2$MnGa and Fe$_2$VAl, (c, d) Fe$_2$Ni$_{1+x}$Ga$_{1-x}$ ($x = 0$ and 0.5), (e) Ni$_{2.5}$Mn$_{0.5}$Ga and Ni$_2$Mn$_{1.5}$Ga$_{0.5}$, (f)~Ni$_2$Mn$_{1+x}$Sn$_{1-x}$ ($x = 0$ and 0.5). Results are presented for different magnetic configurations. In a case of Fe-Ni-Ga, the energy difference is plotted relative to the FM state of the inverse Heusler structure.
} 
\label{figure-3} 
%
\end{figure*}

Table~\ref{table-1} reports the equilibrium lattice constants, bulk moduli, and formation energies calculated for PBE and SCAN. Comparisons with available experimental data and earlier calculations are also listed. 
 Birch-Murnaghan equation of state~\cite{Murnaghan-1944,Birch-1947} was fitted to
 the energy curves to calculate the equilibrium lattice constants. 
 The difference between total energy of the compound and the total energies of corresponding pure elements yields the formation energy  $\Delta E_{\mathrm{form}}$.
SCAN lattice constants are about 1.2~\% smaller while SCAN bulk moduli are about 9 \% larger compared to the corresponding PBE values. 
Regarding the stability of the austenitic phase, SCAN gives negative values of $\Delta E_{\mathrm{form}}$ for all compounds. 
PBE yields positive values of $\Delta E_{\mathrm{form}}$ for three compounds, namely, bcc-NiMn, Ni$_{2}$Mn$_{1.5}$Sn$_{0.5}$, and Fe$_2$Ni$_{1.5}$Ga$_{0.5}$. An overall enhanced stability toward cubic crystal structure is observed in Table~\ref{table-1} when SCAN is compared 
to PBE.

\subsection{Tetragonal distortion}
Possibilities for martensitic transformation (by considering tetragonal distortions of the optimized L2$_1$-cubic structure) are discussed in this subsection. As a matter of fact, 
a martensitic transformation can occur if a tetragonal structure has a lower total energy compared to the cubic structure.
Figure~\ref{figure-3} shows the total energy curves as a function
of the tetragonal $c/a$ ratio for all compounds in the present study. The volume of the supercell was kept constant while the tetragonal distortion was performed in the total energy calculations. The calculated equilibrium $c/a$ ratios and formation energies for tetragonal structures are reported in Table~\ref{table-2}.

\begin{table*}[!htb] 
\caption{The calculated optimized tetragonal $c/a$ ratios and formation energies $\Delta E_{\mathrm{form}}$ (in eV/f.u.) in the  martensitic structure of considered alloys calculated with PBE and SCAN.
}
\begin{ruledtabular}
\begin{tabular}{cc c cc cc cc cc cc cc cc} 
XC Potential & \multicolumn{2}{c}{NiMn}  & \multicolumn{2}{c}{Ni$_2$MnGa} & \multicolumn{2}{c}{Ni$_{2.5}$Mn$_{0.5}$Ga} & \multicolumn{2}{c}{Ni$_2$Mn$_{1.5}$Ga$_{0.5}$} & \multicolumn{2}{c}{Ni$_2$Mn$_{1.5}$Sn$_{0.5}$} & \multicolumn{2}{c}{Fe$_2$Ni$_{1.5}$Ga$_{0.5}$}\\ \hline
             & \textit{c/a} & $\Delta E_{\mathrm{form}}$ & \textit{c/a}  & $\Delta E_{\mathrm{form}}$   & \textit{c/a} &                $\Delta E_{\mathrm{form}}$ & \textit{c/a} & $\Delta E_{\mathrm{form}}$ & \textit{c/a}  & $\Delta E_{\mathrm{form}}$ & \textit{c/a} & $\Delta E_{\mathrm{form}}$  \\ \cline{2-3} \cline{4-5} \cline{6-7} \cline{8-9}  \cline{10-11} \cline{12-13} 

PBE          & 1.4          & -0.058     & 1.25        & -0.632        & 1.23           & -0.491          & 1.35          & -0.234         & 1.3            & -0.710           & 1.45           & 0.000            \\
SCAN         & 1.3          & -0.519     & 1.2         & -1.636        & 1.25           & -1.394          & -             & -              & 1.15           & -0.708           & 1.45           & -0.742          
\end{tabular}
\end{ruledtabular}
\label{table-2}
\end{table*}

\begin{table*}[!htb] 
\caption{The calculated reference state, $c/a$ ratio, total and partial magnetic moments ($\mu_{tot}$ and $\mu_\mathrm{Z}$) for stable austenite and martensite phases of the considered compounds, which were computed with  PBE and SCAN. 
}
\begin{ruledtabular}
\begin{tabular}{cc c cc cc cc cc cc cc cc}
System& & \multicolumn{2} {c}{Ref. state}  &\multicolumn{2} {c}{$c/a$ ratio} & \multicolumn{2} {c}{$\mu_{\rm{Ni}}$}&\multicolumn{2} {c}{$\mu_{\rm{Mn_1}}$}& \multicolumn{2} {c}{$\mu_{\rm{Mn_2}}$} & \multicolumn{2} {c}{$\mu_{\rm{Fe}}$} & \multicolumn{2} {c}{$\mu_{tot}$} \\
 &    & PBE & SCAN &  PBE& SCAN & PBE & SCAN & PBE &SCAN & PBE & SCAN & PBE & SCAN  & PBE & SCAN\\ \hline
\multirow{2}{*}{NiMn}    & aust.   & FM  & FM &  1.0  & 1.0 & 0.816 & 0.978 &  3.523 & 3.781  & & & & &4.339 & 4.758 \\ 
                         & mart.  & AFM-1&AFM-1& 1.4  & 1.3 & 0.0   & 0.0   &  3.292 & 3.737 & -3.292 & -3.737 & & & 0.0 & 0.0 \\
\multirow{2}{*}{Ni$_2$MnGa}& aust. & FM & FM &  1.0  & 1.0 & 0.366 & 0.539 &  3.401 & 3.689  & & & & &4.082 & 4.72 \\ 
                           & mart. & FM & FM &  1.25 & 1.2 & 0.437 & 0.548  &  3.321 & 3.634 &  &   & & & 4.134 & 4.667 \\                         

\multirow{2}{*}{Ni$_{2.5}$Mn$_{0.5}$Ga}& aust. & FM & FM &  1.0  & 1.0  & 0.26 & 0.376  &  3.393 & 3.647   & & & & &2.324 & 2.737   \\ 
                                       & mart. & FM & FM &  1.23 & 1.25 & 0.29 & 0.353  &   3.376 &  3.59  &  &   & & & 2.383&2.636   \\  

\multirow{2}{*}{Ni$_2$Mn$_{1.5}$Ga$_{0.5}$}& aust. & FIM & FM &  1.0  & 1.0 & 0.24 & 0.779 &  3.395 & 3.706  & -3.497& 3.798& & &2.105 & 7.181 \\ 
                                            & mart. & FIM & -  &  1.35 &  - & 0.171&  - & 3.243 &   - & -3.36 & -  & & & 1.875    & -    \\  
\multirow{2}{*}{Ni$_2$MnSn}& aust. & FM & FM &  1.0  & 1.0 & 0.248 & 1.028 &  3.59 & 4.16  & & & & &4.036 & 4.450 \\ 
                           & mart. &  - & -  &     - &   - &   -   &   -   &  -     &   - &  &   & & & -    & -    \\     
\multirow{2}{*}{Ni$_2$Mn$_{1.5}$Sn$_{0.5}$}& aust. & FIM & FIM &  1.0  & 1.0 & 0.152 & 0.644 &  3.481 & 3.84  & -3.672& 3.927& & &1.922 & 7.065 \\ 
                                           & mart. &  FIM &  FM &   1.3 & 1.15&   0.124& 0.682& 3.366 & 3.801 & -3.593&3.901 & & & 1.784& 7.082    \\
\multirow{2}{*}{Fe$_2$NiGa}& aust. & FM & FM &  1.0  & 1.0 & 0.495 & 0.533 &   &   & & & 2.275& 2.444& 4.984 & 5.331 \\ 
                           & mart. &  - & -  &     - &   - &   -   &   -   &     &    &  &   & - & - & -    & -    \\    
\multirow{2}{*}{Fe$_2$Ni$_{1.5}$Ga$_{0.5}$}& aust. & FM & FM &  1.0  & 1.0 & 0.603 & 0.716 &   &   & & & 2.434& 2.703&  5.731& 6.427 \\ 
                           & mart. & FM & FM  &  1.45 &   1.45 &    0.616 & 0.673 & &  &   &   & 2.599 & 2.86 & 6.059   & 6.656    \\                                                                         
\end{tabular}
\end{ruledtabular}
\label{table-3}
\end{table*}

We start by considering the binary compound NiMn, which is an antiferromagnet~\cite{Kasper-1959,Entel-2011} with CsCl structure and has a N\'eel temperature higher than 1000 K. This compound undergoes a structural phase transformation from the bcc-like austenite ($\beta$-NiMn) to the L1$_0$-tetragonal martensite ($\alpha$-NiMn with fcc-like structure) at a high temperature of about 1000 K during cooling. As shown in Fig.~\ref{figure-3}(a), a crossover from the FM bcc-structure ($c/a = 1$) to the AFM L1$_0$ (fcc-like) structure is obtained both with PBE ($c/a = 1.4$) and SCAN ($c/a = 1.3$). 
Curiously, for the FM state, SCAN produces a slight cubic symmetry breaking since the $c/a$ minimum is shifted from 1 to 1.05.
PBE gives the observed $\alpha$ phase with fcc structure and provides a good estimate for martensitic transformation temperature~\cite{Entel-2011,temperature} while SCAN degrades the agreement with experiment concerning the $c/a$ ratio and the martensitic transformation temperature.  
 Moreover, both PBE and SCAN yield the AFM-1 layered structure of the martensite phase as energetically more stable. Similar trends of the total energy curves plotted in Fig.~\ref{figure-3}(a)
 have been reported by Godlevsky and Rabe~\cite{Godlevsky-2001} (LDA) and Entel \textit{et al}.\cite{Entel-2011} (GGA) indicating that corrections beyond LDA are not too strong.
 

\begin{figure*}[!htb] 
\centerline{
\includegraphics[width=7.5cm]{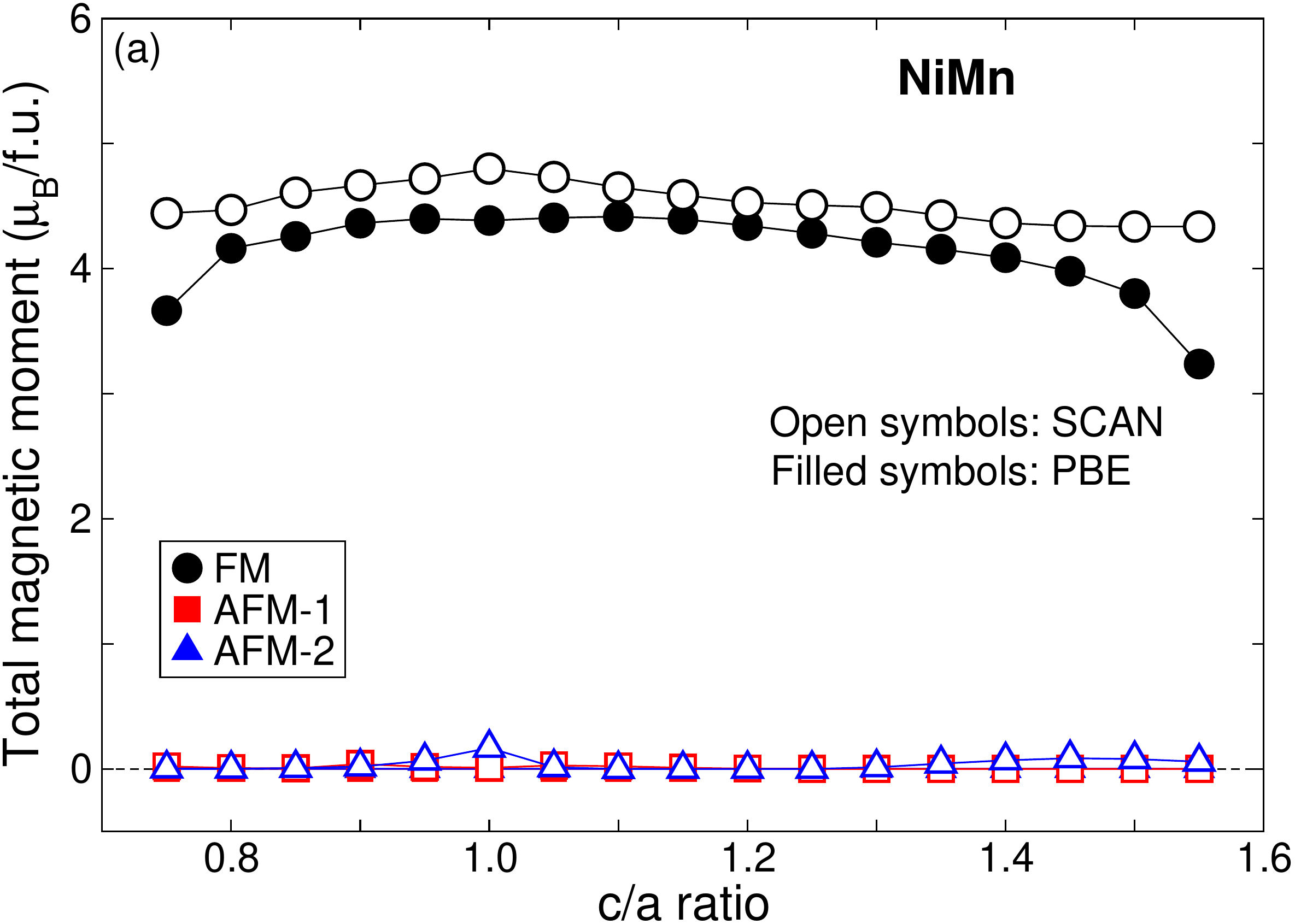}
 \hspace*{0.5cm}
\includegraphics[width=7.5cm]{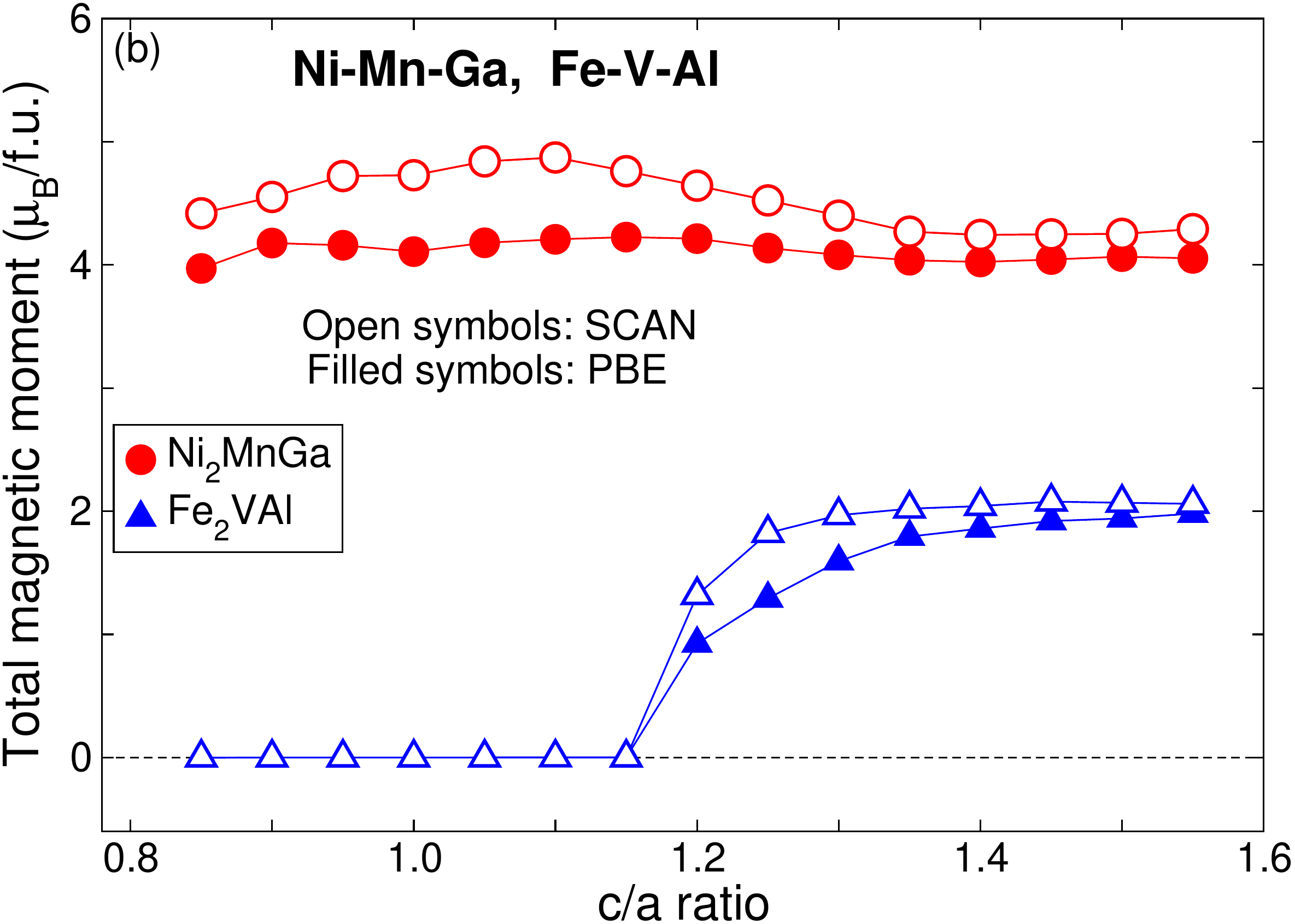}
           }   
           \centerline{
\includegraphics[width=7.5cm]{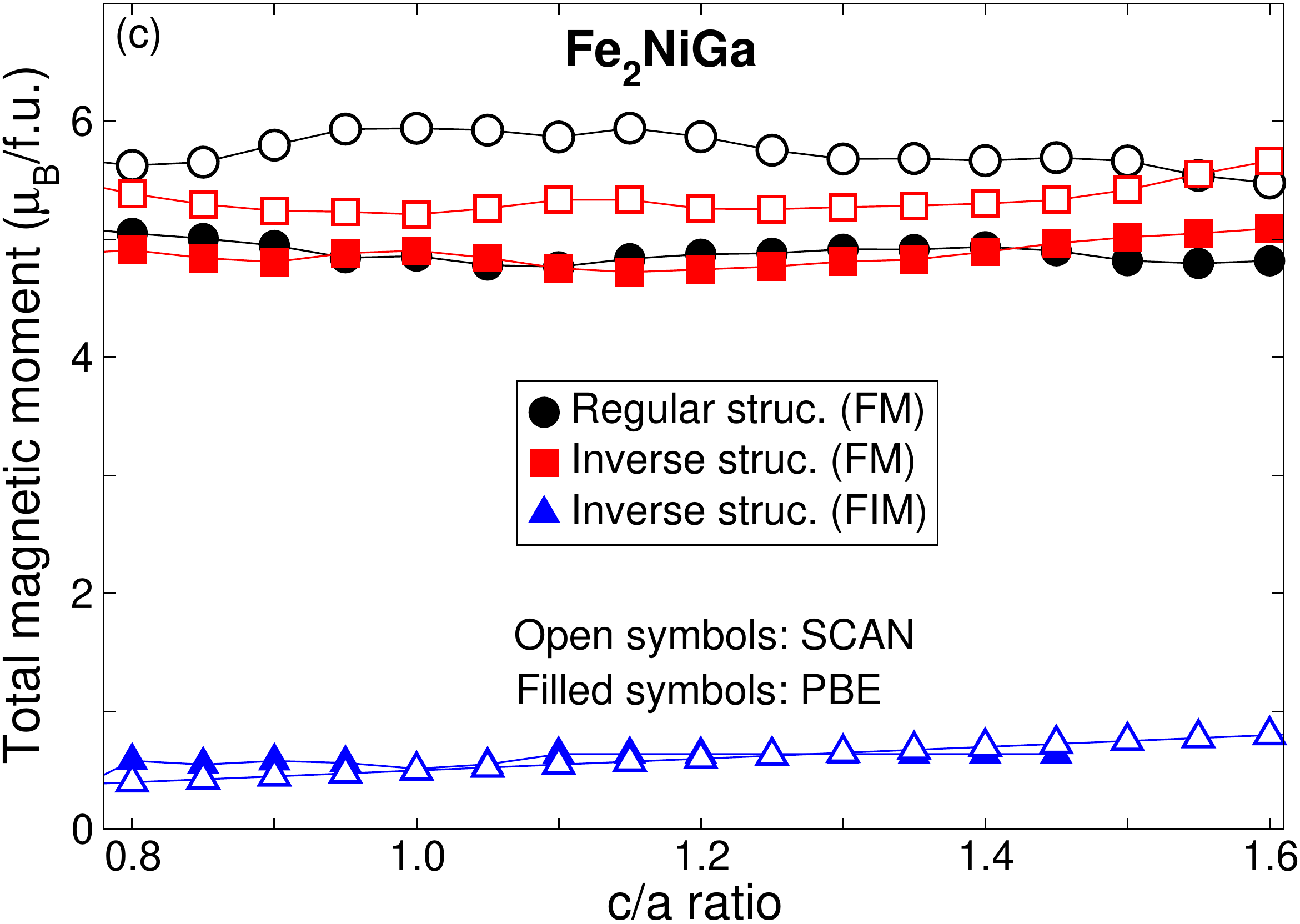}
 \hspace*{0.5cm}
\includegraphics[width=7.5cm]{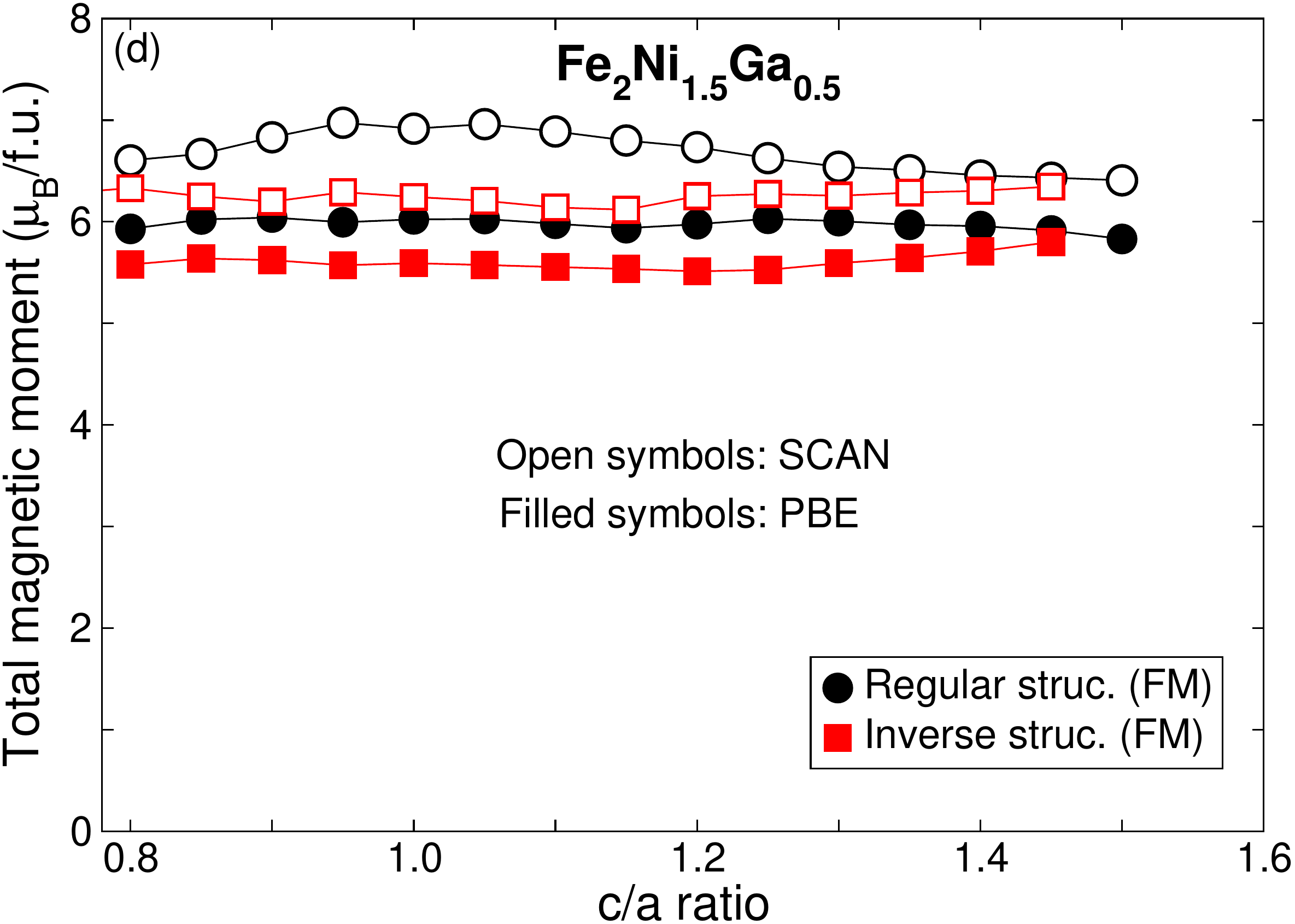}
           }  
           \centerline{
\includegraphics[width=7.5cm]{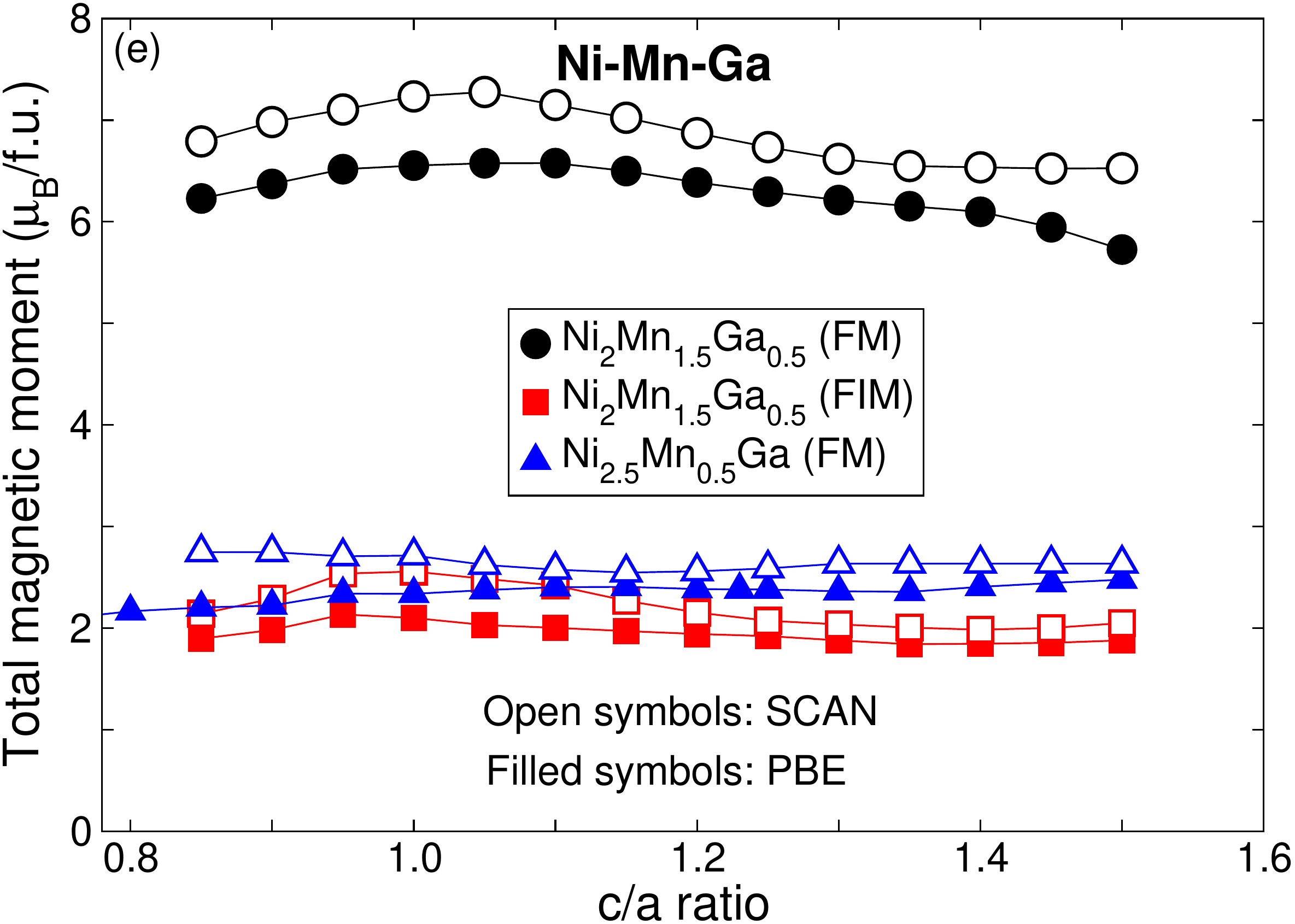}
 \hspace*{0.5cm}
\includegraphics[width=7.5cm]{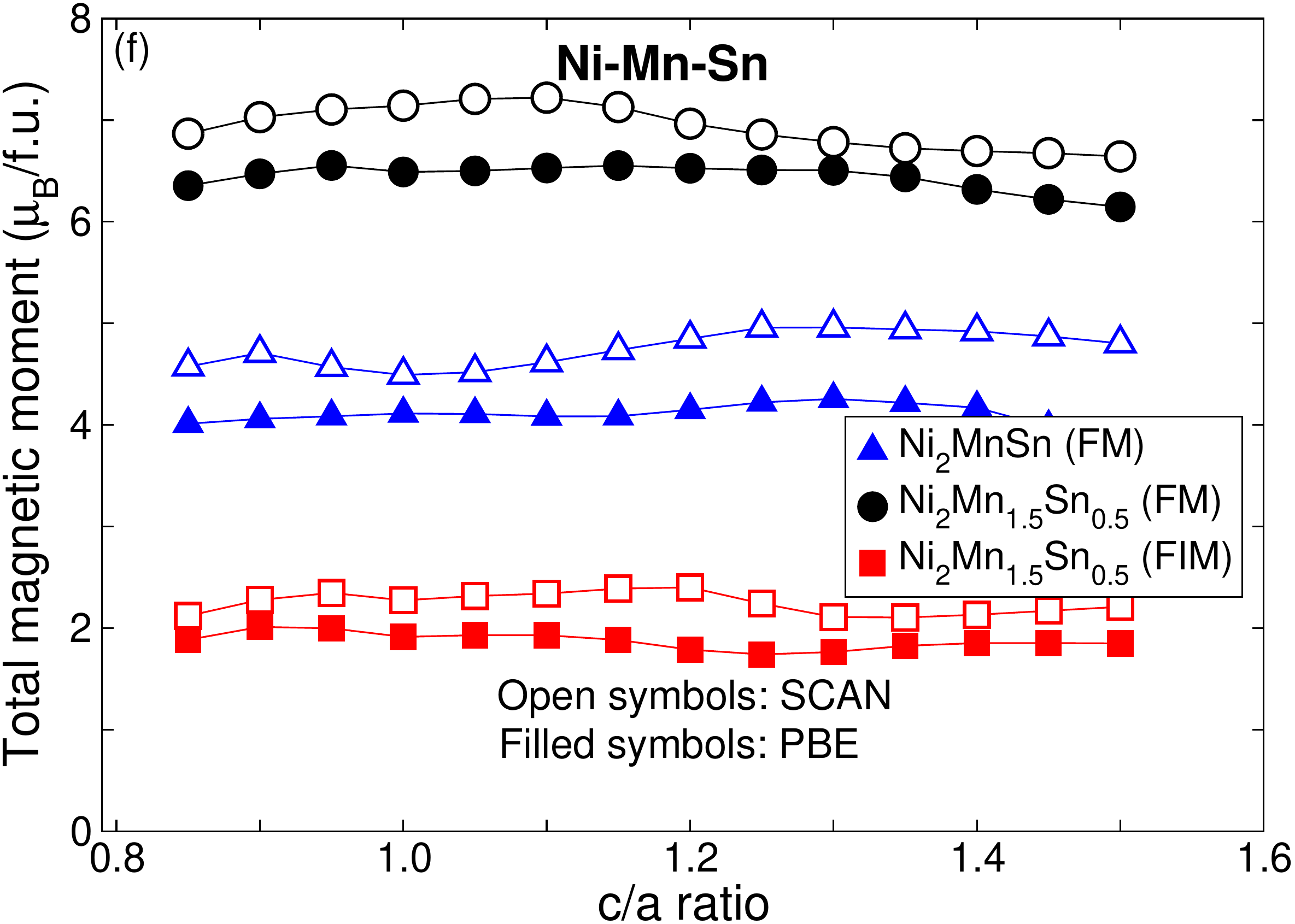}
           } 
\caption{(Color online) 
 The calculated total magnetic moments as a function of tetragonal ratio $c/a$ for (a)~NiMn, (b)
 Ni$_2$MnGa and Fe$_2$VAl, (c, d) Fe$_2$Ni$_{1+x}$Ga$_{1-x}$ ($x = 0$ and 0.5), (e) Ni$_{2.5}$Mn$_{0.5}$Ga and Ni$_2$Mn$_{1.5}$Ga$_{0.5}$, (f)
Ni$_2$Mn$_{1+x}$Sn$_{1-x}$ ($x = 0$ and 0.5). Results are presented for different magnetic configurations. 
} 
\label{figure-4} 
%
\end{figure*}

Figures~\ref{figure-3}(b)-(f) illustrate the comparison of PBE and SCAN for a series of FM Ni$_{2+x}$Mn$_{1-x}$Ga and Fe$_2$Ni$_{1+x}$Ga$_{1-x}$,
 FIM Ni$_2$Mn$_{1+x}$(Ga, Sn)$_{1-x}$, and non-magnetic Fe$_2$VAl.
PBE and SCAN cubic structures are found to be stable only for stoichiometric Fe$_2$VAl (Fig.~\ref{figure-3}(b)), Fe$_2$NiGa (Fig.~\ref{figure-3}(c)), and  Ni$_2$MnSn (Fig.~\ref{figure-3}(f)) preventing martensitic
transitions. 

Concerning Ni$_2$MnGa, Fig.~\ref{figure-3}(b) shows that PBE gives a local minima at $c/a = 1$ and a global minima $c/a = 1.25$ in agreement with earlier calculations \cite{Entel-2006,Entel-2011,Kart-2008} while SCAN yields the local minima at $c/a = 0.95$ and the global minima at $c/a = 1.2$. Interestingly, SCAN $c/a$  is in excellent agreement with experimental~\cite{Martynov-1992,Sozinov-2002} $c/a~=~1.18~\pm~0.02$.
Moreover, SCAN leads to a larger energy difference between the metastable austenite and martensite phase with respect to PBE. This implies that the predicted temperature of martensitic transformation from SCAN ($T_m \approx$ 153 K) is closer to the experimental value \cite{Webster-1984} ($T_m \approx$~202~K) with respect to PBE ($T_m \approx$ 107 K) \cite{temperature}. 

For Fe$_2$Ni$_{1.5}$Ga$_{0.5}$, according to Fig.~\ref{figure-3}(d) and $\Delta E_{\mathrm{form}}$ in Tables~\ref{table-1} and ~\ref{table-2}, a
structural transition  from FM austenite with inverse Heusler structure to FM martensite with regular structure is predicted 
 only within SCAN. PBE yields unstable martensitic phase with regular structure due to zero formation energy. The overall behavior of the total energy curves is similar for PBE and SCAN.

In the case of  Ni$_{2.5}$Mn$_{0.5}$Ga 
(Fig.~\ref{figure-3}(e)), PBE and SCAN total energy curves almost conincide and predict a martensitic phase with $c/a$ about 1.25 while the austenitic phase is not favorable. However, the stability of this compound has been questioned by experiments~\cite{Khovailo-2005}.

Surprisingly, for Mn-excess Ni$_2$Mn$_{1.5}$(Ga, Sn)$_{0.5}$ (Figs.~\ref{figure-3}(e, f)),
SCAN disagrees with PBE by stabilizing FM instead of FIM ground state in contrast with PBE and SCAN agreement for the richest Mn-excess Ni$_2$Mn$_{1+x}$Z$_{1-x}$ with AFM-1 ground state as shown in Fig.~\ref{figure-3}(a). Moreover, 
 although PBE and SCAN total energy curves as a function of $c/a$ have a similar behavior for Ni$_2$Mn(Ga, Sn), they significantly disagree for Ni$_2$Mn$_{1.5}$(Ga, Sn)$_{0.5}$. In particular, 
PBE predicts for both systems
 the austenite-martensite transformation in the FIM state in agreement with earlier calculations\cite{Ye-2010,Entel-2014,Xiao-2012,Xiao-2014}. PBE $c/a$ global minima for Ni$_2$Mn$_{1.5}$Ga$_{0.5}$ and Ni$_2$Mn$_{1.5}$Sn$_{0.5}$ are 1.35 and 1.3, respectively.  
Experimental $c/a$ ratios~\cite{Cakir-2013,Cakir-2015} of about 1.28 and 1.24 for non-modulated L1$_0$-tetragonal structure of Ni$_2$Mn$_{1.52}$Ga$_{0.48}$ and Ni$_2$Mn$_{1.52}$Sn$_{0.48}$, respectively,
 have been reported.
Regarding SCAN, global minima are observed at $c/a \approx 1.25$ for Ni$_2$Mn$_{1.5}$Ga$_{0.5}$ and at  $c/a \approx 1.15$ for Ni$_2$Mn$_{1.5}$Sn$_{0.5}$ in FM state.
 Nevertheless, experiments~\cite{Aksoy-2009,Cakir-2013,Cakir-2015} suggest 
that L1$_0$-tetragonal phases for Mn-excess Ni-Mn-(Ga, Sn) yield almost degenerate AFM and FM ground states. Therefore, it is difficult to conclude if either PBE or SCAN give a better agreement with experiment.


\subsection{Magnetic moments}
Table~\ref{table-3} displays the value of total and partial magnetic moments for both austenite and martensite phases calculated with PBE and SCAN. 
As a general trend, SCAN gives higher magnetic moments as compared to PBE values. 
\begin{figure*}[!htb] 
\centerline{
\includegraphics[width=7.5cm]{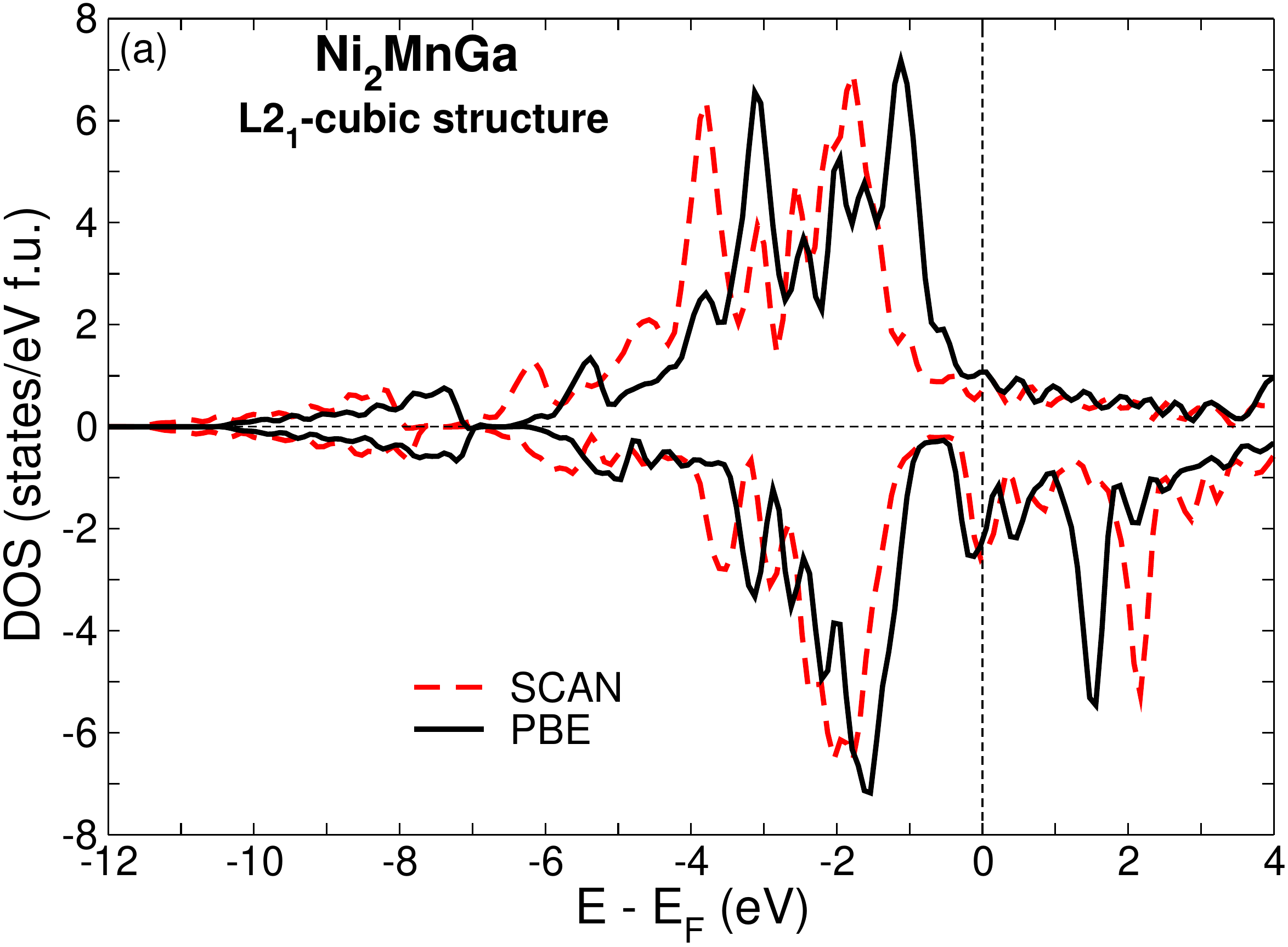}
 \hspace*{0.5cm}
\includegraphics[width=7.5cm]{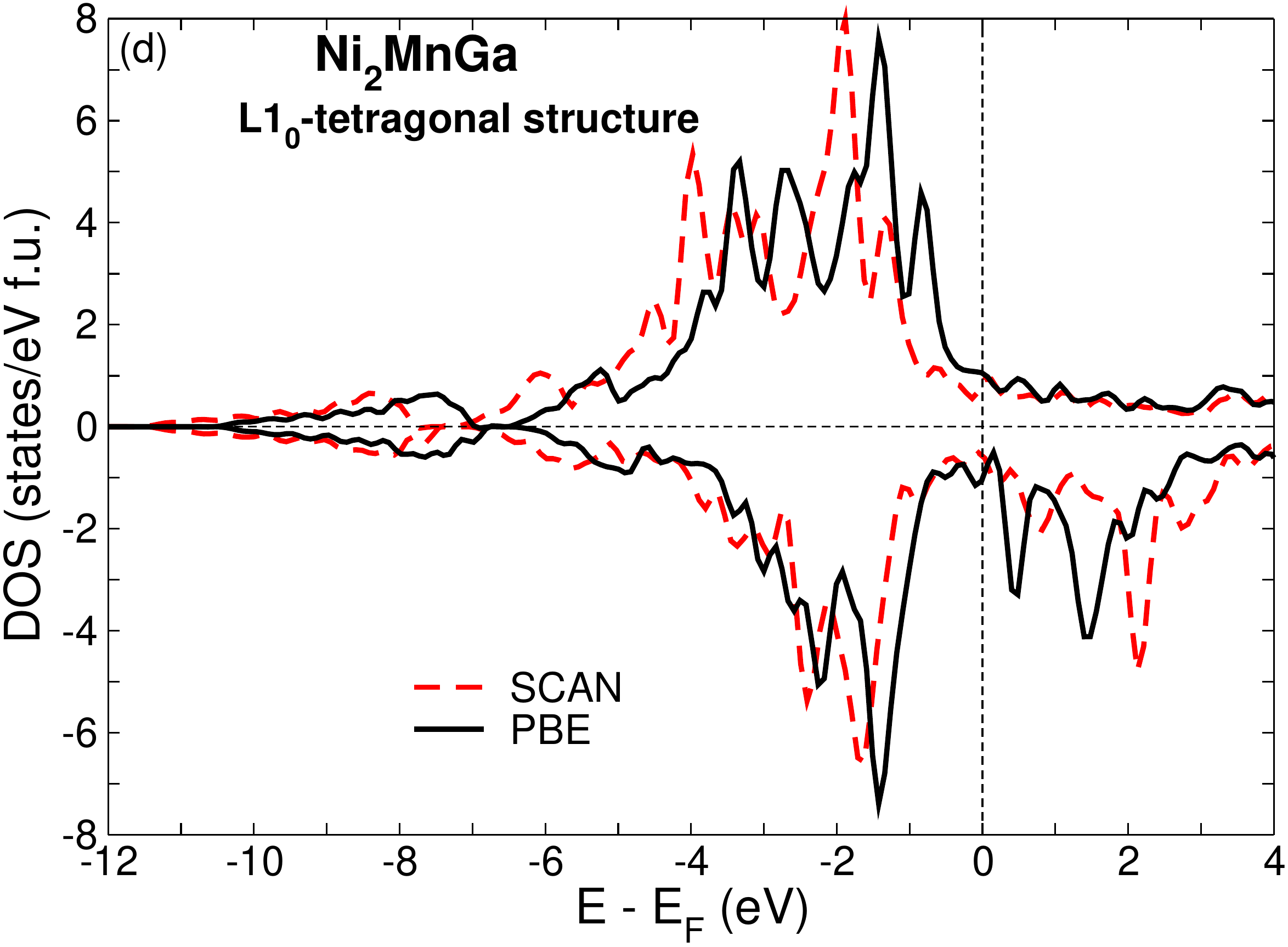}
           }   
           \centerline{
\includegraphics[width=7.5cm]{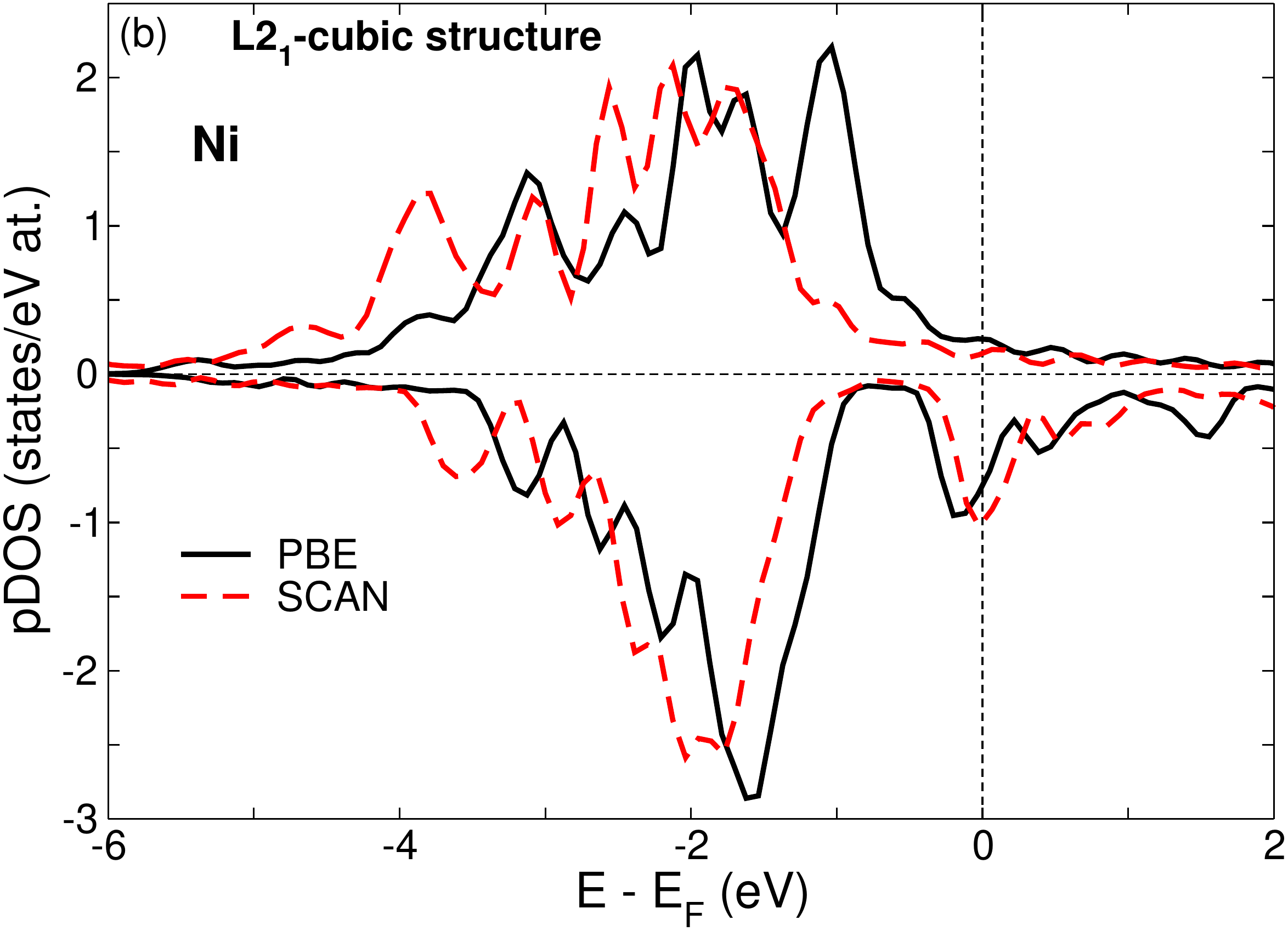}
 \hspace*{0.5cm}
\includegraphics[width=7.5cm]{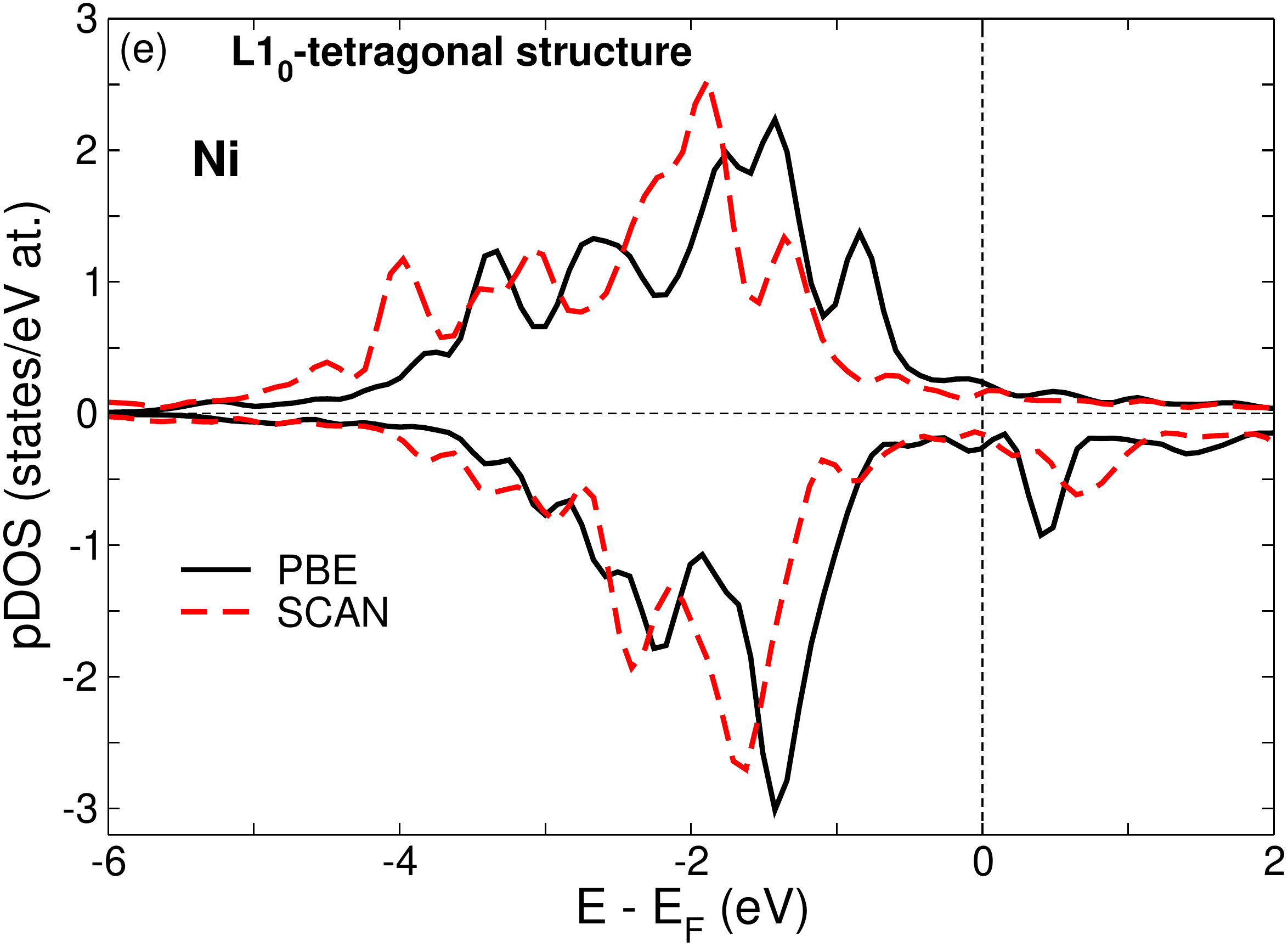}
           } 
  \centerline{
           \includegraphics[width=7.5cm]{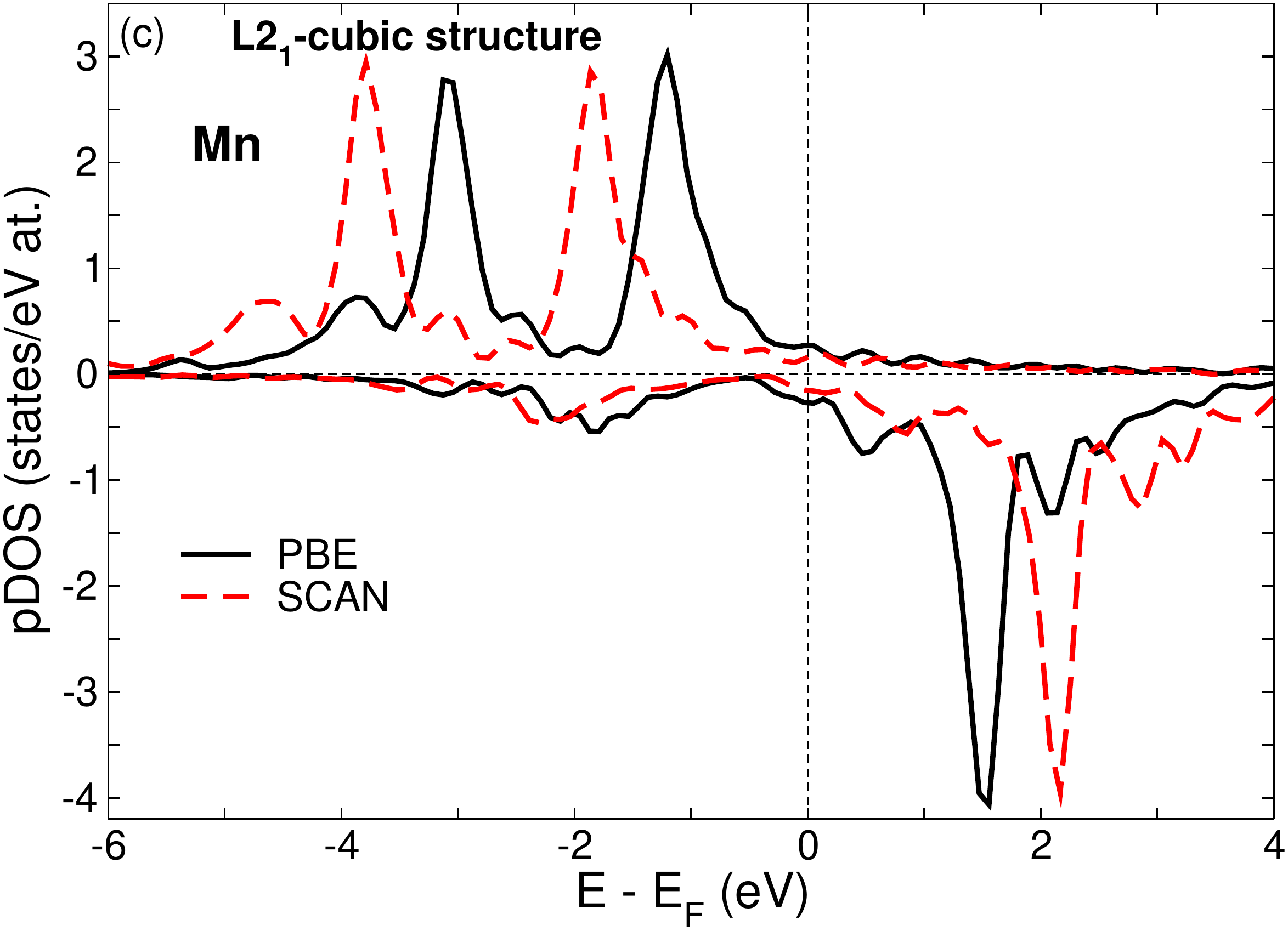}
 \hspace*{0.5cm}
\includegraphics[width=7.5cm]{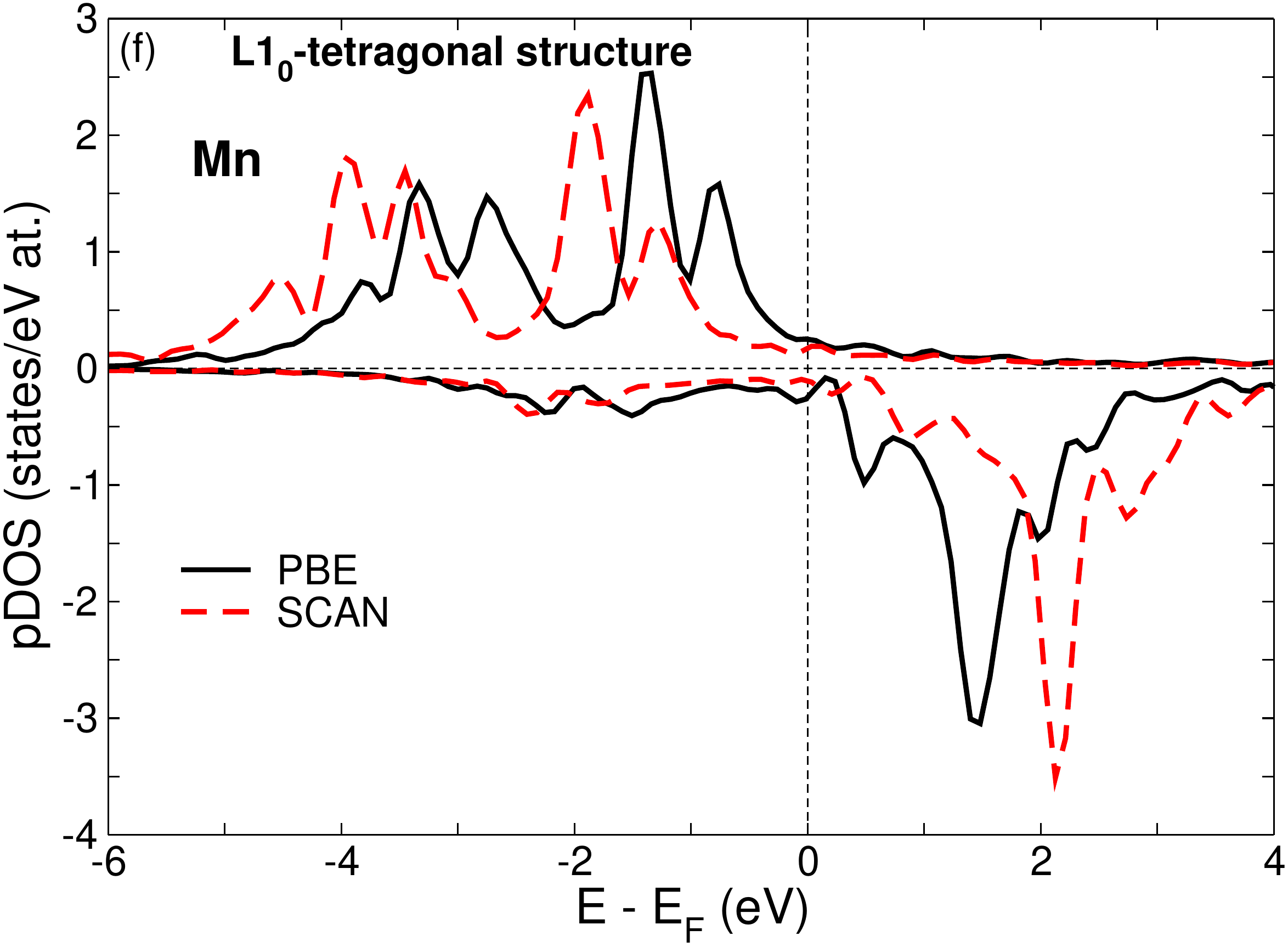}
           } 
\caption{(Color online) 
 The total DOS and 3$d$ Ni and 3$d$ Mn partial DOS calculated with PBE and SCAN for (a, b, c) austenite and (d, e, f) martensite phases of Ni$_2$MnGa alloy.
} 
\label{figure-5} 
%
\end{figure*}

\begin{figure*}[!htb] 
\centerline{
\includegraphics[width=7.5cm]{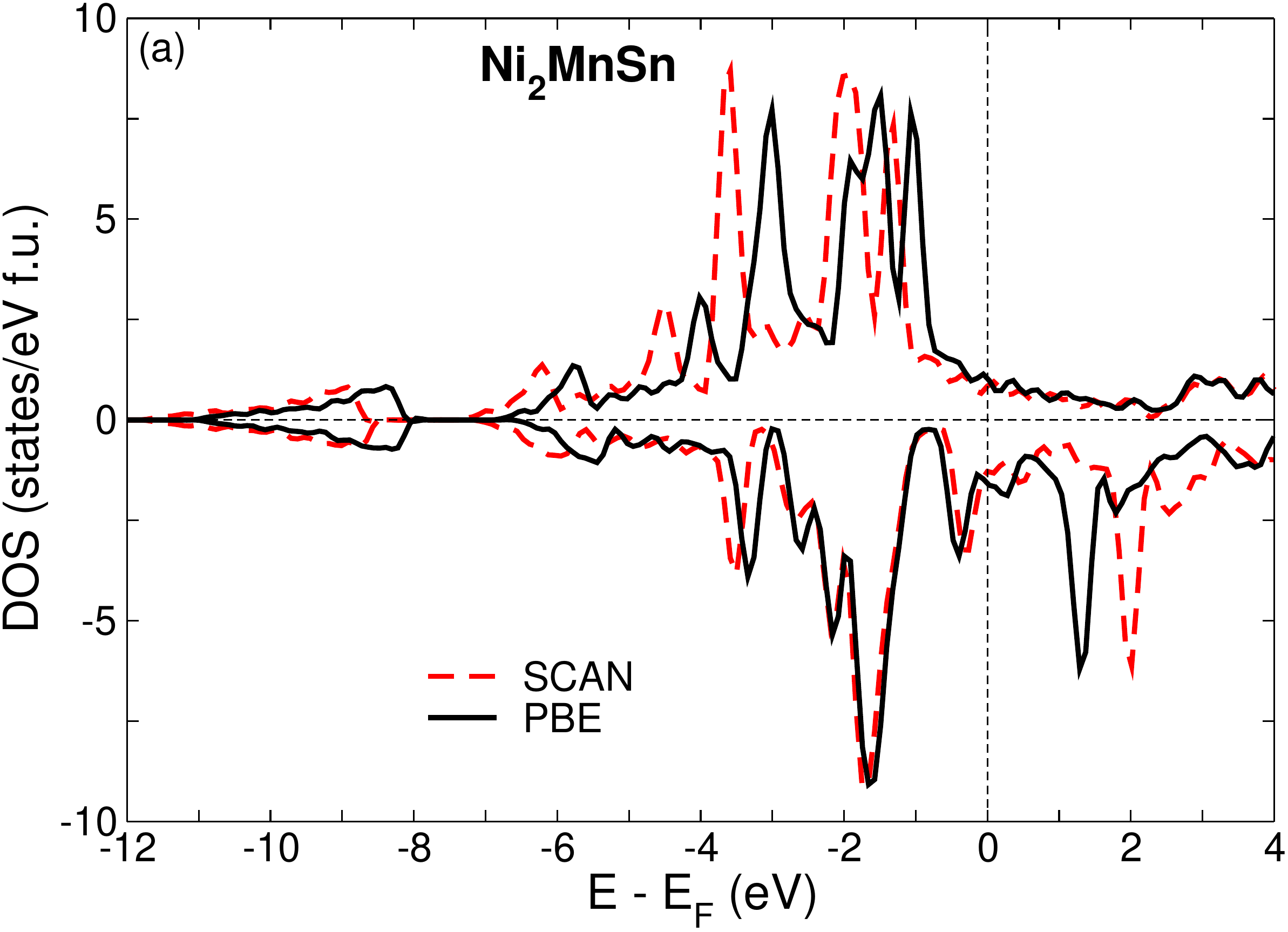}
 \hspace*{0.5cm}
\includegraphics[width=7.5cm]{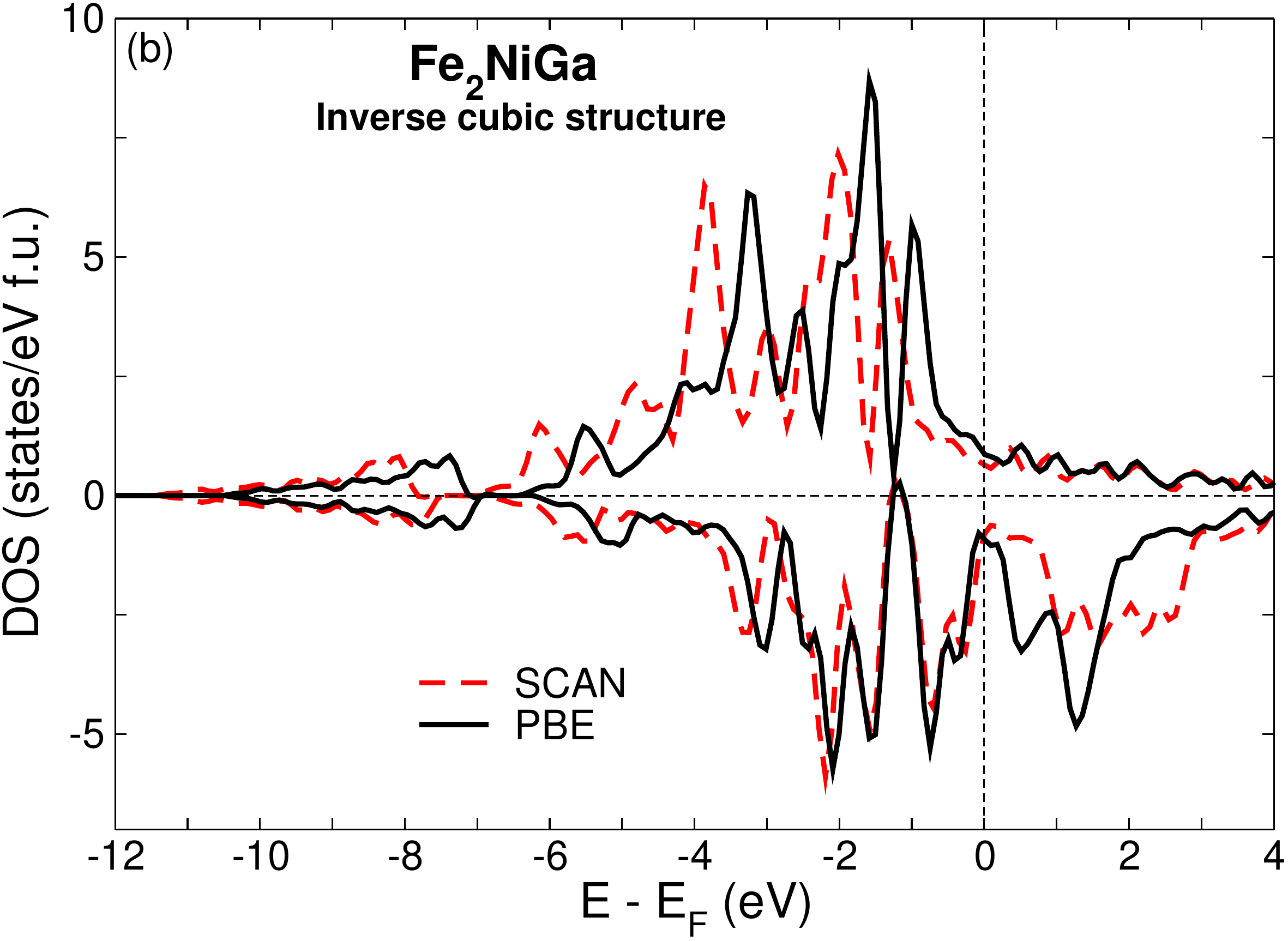}
           }   
           \centerline{
\includegraphics[width=7.5cm]{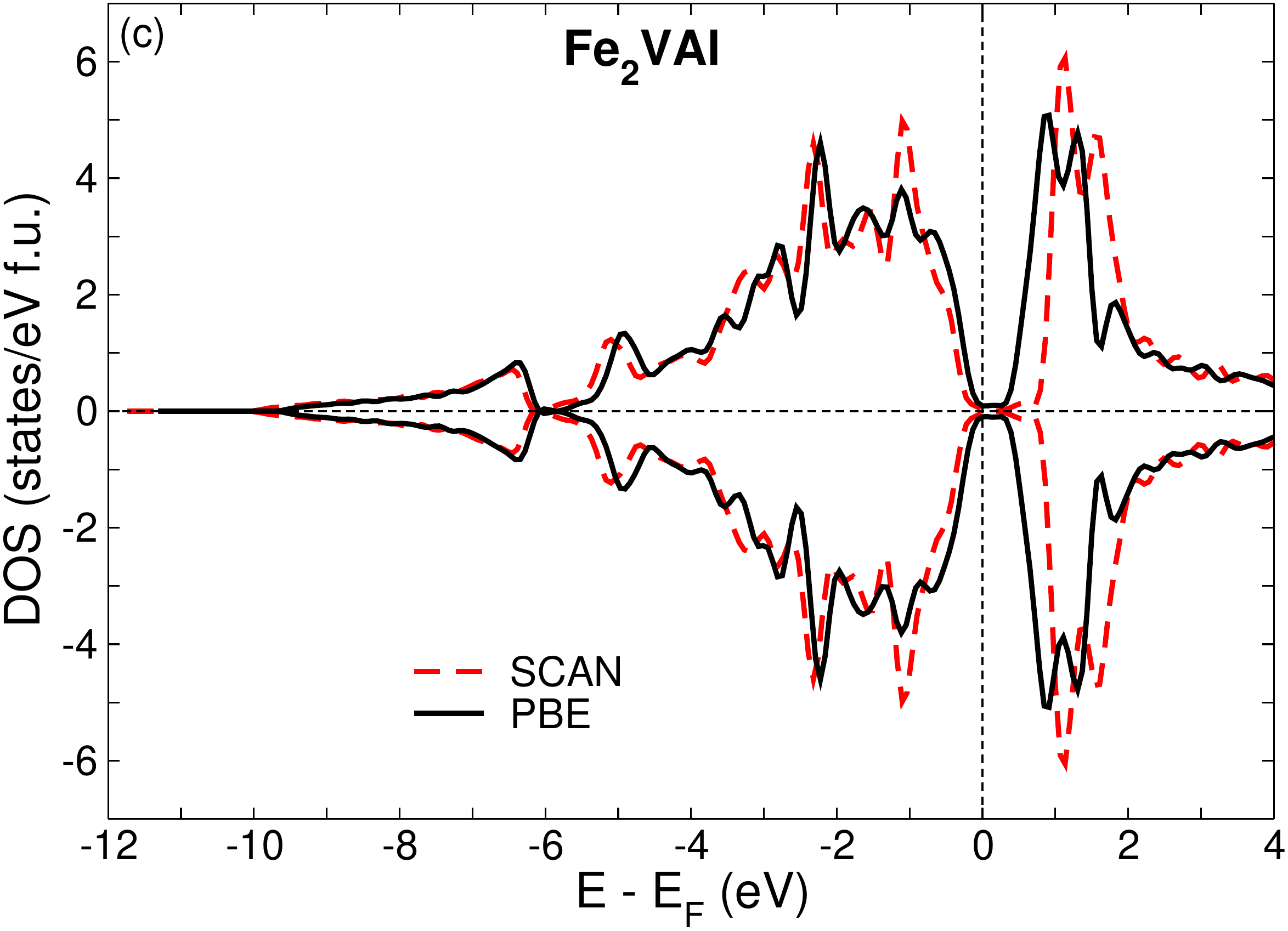}
           }   
\caption{(Color online) 
 The total DOS calculated with PBE and SCAN for the austenitic phase of (a) Ni$_2$MnSn, (b) Fe$_2$NiGa, and (c) Fe$_2$VAl Heusler compounds. Here the solid and dash lines denote the PBE and SCAN calculations.
} 
\label{figure-6} 
%
\end{figure*}

Figure 4 shows the total magnetic moments for FM, FIM, AFM-1, and AFM-2 configurations as a function of tetragonal deformation $c/a$. Mn and Fe atoms provide the largest contribution to the total magnetic moment.     
Most of the curves in Fig. 4 display a gentle behavior as a function of $c/a$ with the exception of meta-stable FM martensitic phase ($c/a > 1.15$) of Fe$_2$VAl. Fe$_2$VAl is non-magnetic in the range of $0.8 \leq c/a < 1.15$, but it becomes FM when $c/a > 1.15$ for PBE and SCAN.
However, the martensitic phase is not stable for Fe$_2$VAl (see Fig.~\ref{figure-3}(b)). Generally,
in the considered $c/a$ range, the difference between SCAN and PBE for the magnetic moments is about 10 \%. 
As illustrated in Fig.~\ref{figure-4}(b), according to PBE, Ni$_2$MnGa has a higher magnetic moment for martensite ($4.134$ $\mu_B$/f.u.) than that for austenite ($4.082$~$\mu_B$/f.u.), while SCAN gives a slightly lower magnetic moment for martensite ($4.67$ $\mu_B$/f.u.) in comparison with austenite ($4.72$ $\mu_B$/f.u.). 
Experiments~\cite{Webster-1984} find  a drop in the magnetization across the martensitic transformation from FM martensite to FM austenite (for magnetic fields higher than 0.8 T upon heating) suggesting that the low-temperature tetragonal phase has a higher magnetic moment compared to austenite as predicted by PBE. 


\subsection{Electronic structure}

In this subsection we discuss correlation effects on the electronic structure of Ni$_2$Mn(Ga, Sn), Fe$_2$NiGa, and Fe$_2$VAl.
Figure~\ref{figure-5} shows the spin-resolved total density of states (DOS) and partial DOS (pDOS) for $d$-orbitals for Ni$_2$MnGa in the L2$_1$-cubic and L1$_0$-tetragonal phase. The DOS and pDOSs calculated with PBE reproduce features already investigated earlier \cite{Ayuela-2002,Entel-2006,Kart-2008}.
For instance, Ga 4$s$ electrons are responsible for the contribution to the total DOS in the  valence band below -7~eV.  
The upper bands below and above the Fermi energy ($E_F$) are due to 3$d$ electrons of Mn and Ni. The majority $d$ states of Mn hybridizes with the spin up $d$ states of Ni while the minority Mn $d$
states are located above $E_F$. 
The tetragonal distortion ($c/a > 1$) changes slightly the electronic structure and leads to the splitting of the peak in the minority 3$d$ states of the Ni band near $E_F$ into two parts.
While one part is shifted above the Fermi level (and therefore is not occupied any more), the other part is shifted to a lower energy.
Thereby the band energy of the tetragonal structure is lower than the band energy of the austenite structure.
In general, the structural instability and formation of martensite can be associated with a Jahn-Teller band scenario involving mostly contribution of Ni 3$d$ states~\cite{Ayuela-2002,Brown-1999,Entel-2006,Opeil-2008} as illustrated in Fig.~\ref{figure-5}. 
SCAN preserves the basic features of the DOS.
It produces the exchange splitting by about 0.5 eV both in the austenite and martensite. 

Figure~\ref{figure-6} shows the total DOS calculated with PBE and SCAN for Ni$_2$MnSn, Fe$_2$VAl, and Fe$_2$NiGa at their equilibrium volume in the austenitic phase. 
As in the case of Ni$_2$MnGa, we find that the total DOS curves produced by SCAN are modified by exchange splitting enhancement with respect to PBE. 
 For Fe$_2$VAl, the vanishing magnetic moment (for PBE and SCAN) agrees with the Slater-Pauling rule\cite{Slater} and the DOS has no exchange splitting implying that the material is a  nonmagnetic semi-metal with a
very small DOS at $E_F$. 
Moreover, SCAN enlarges the corresponding pseudogap.
Since, PBE and SCAN DOSs are very similar for the occupied states,
the corrections beyond GGA are smaller for non-magnetic compared to magnetic Heusler compounds.
To better understand the SCAN effects of correlation on the DOS, one can also considered more correlated Heusler compounds, for which GGA fails.  The Co$_2$FeSi is one of such a material since GGA fails to produce a half metallic gap, while the GW method~\cite{Meinert-2012} reproduces experimental magnetic moment and half metallic energy gap. Interestingly, smaller GW band-gap corrections are found for quaternary Heusler like (CoFe)TiAl~\cite{Tas-2016}. We show in Fig.~S3 of the supplementary materials (SM)~\cite{SM} that SCAN reproduce very well the GW corrections for the DOS of these materials (we also show total energy results in Figs. S1 and S2 of the SM).

\section{Conclusions}
In the present work, the structural, magnetic and electronic properties of a series of Heusler alloys were investigated in the framework of DFT calculations. SCAN can be viewed as a correction of PBE containing extra semilocal information. Therefore, SCAN corrections for correlation effects play an important role in determining exchange interactions as in the case of DFT + $U$ methods studied by different authors~\cite{Himmetoglu-2012,Hasnip-2013} and half metallic energy gaps in the DOS as in the case of the GW approach~\cite{Meinert-2012, Tas-2016}.

The present investigation suggests that corrections beyond GGA are rather minor for FM  Ni$_{2+x}$Mn$_{1-x}$Ga, Fe$_2$Ni$_{1+x}$Ga$_{1-x}$ and non-magnetic Fe$_2$VAl compounds. However,
significant differences between PBE and SCAN are observed for Mn-excess compounds such as Ni$_2$Mn$_{1+x}$(Ga, Sn)$_{1-x}$, where localized magnetic moments on Mn atoms couple via oscillating RKKY interactions. Thus, the magnetic behavior of these compounds is very sensitive to the distance between Mn atoms~\cite{Ye-2010}.
According to experiments~\cite{Aksoy-2009} in compounds with Mn excess atoms the total magnetic moment decreases at the austenite-martensite transformation on cooling due to the smaller neighbor Mn-Mn distance, which becomes antiferromagnetically coupled. Regarding simulations, SCAN tends to favor FM solutions in austenite and martensite while PBE yields FIM ground state. Therefore, we conclude that the present corrections beyond GGA could be exaggerated for Mn-Mn FM interactions at short distances.

Pseudogap, spin density and charge density waves (commensurate or nor commensurate) might change the solutions landscape in Heusler alloys as observed by various authors~\cite{Opeil-2008,Dutta-2016,Ye-2010,Kundu-2017}, some of these spin wave or charge density instabilities could be driven by Fermi surface nestings \cite{Dugdale-2006}. Interestingly, SCAN simulations for YBa$_2$Cu$_3$O$_7$, give many solutions almost degenerate with the ground state in the so-called pseudogap regime~\cite{Zhang-2018}. 
Nevertheless, some issues related to exaggerated FM coupling in SCAN  remain~\cite{Isaacs-2018,Ekholm-2018,Fu-2018} regardless the   supercell size in the calculations
and should be addressed in improved versions of this functional.

\section{Acknowledgments}
This work was supported by RSF-Russian Science Foundation
No. 17-72-20022. Calculations for Fe$_2$NiGa were supported by RSF No. 18-12-00283. B.B. acknowledges support from the COST Action CA16218.




\begin{thebibliography}{46}
\bibitem{temperature}
 To estimate $T_m$ temperature, the crude approximation in the form of $\Delta E \approx k_BT_m$ was considered. Here $k_B$ is the Boltzman constant and the conversion factor 1 meV/atom = 11.6 K was used.
\bibitem{Slater}
The Slater-Pauling rule is as follows: $\mu_{tot}$/f.u. = $z$ - 24, here $z$ is the number of valence electrons. Fe$_2$VAl has 24 valence electrons.
\end{thebibliography}
\end{document}